FAKULTÄT
FÜR !NFORMATIK
Faculty of Informatics

# CERES in aussagenlogischen Beweisschemata

MASTERARBEIT

zur Erlangung des akademischen Grades

**Master of Science**

im Rahmen des Studiums

**Computational Logic**

eingereicht von

**Andrea Condoluci**
Matrikelnummer 1528762

an der Fakultät für Informatik
der Technischen Universität Wien

Betreuung: Univ.Prof. Dr.phil. Alexander Leitsch

Wien, 7. Oktober 2016

| Andrea Condoluci | Alexander Leitsch |
|---|---|

FAKULTÄT FÜR !NFORMATIK
Faculty of Informatics

# CERES in Propositional Proof Schemata

## MASTER'S THESIS

submitted in partial fulfillment of the requirements for the degree of

**Master of Science**

in

**Computational Logic**

by

**Andrea Condoluci**
Registration Number 1528762

to the Faculty of Informatics

at the TU Wien

Advisor: Univ.Prof. Dr.phil. Alexander Leitsch

Vienna, 7th October, 2016

| Andrea Condoluci | Alexander Leitsch |
|---|---|

# Erklärung zur Verfassung der Arbeit

Andrea Condoluci
Favoritenstraße 9-11, 1040 Wien, Austria

Hiermit erkläre ich, dass ich diese Arbeit selbständig verfasst habe, dass ich die verwendeten Quellen und Hilfsmittel vollständig angegeben habe und dass ich die Stellen der Arbeit – einschließlich Tabellen, Karten und Abbildungen –, die anderen Werken oder dem Internet im Wortlaut oder dem Sinn nach entnommen sind, auf jeden Fall unter Angabe der Quelle als Entlehnung kenntlich gemacht habe.

Wien, 7. Oktober 2016

Andrea Condoluci

# Acknowledgements

This thesis is the conclusion of my Master's studies, therefore these acknowledgements apply not only to the hundred pages of the following scientific text, but also to the last seven hundred astounding days of my life.

Firstly, my gratitude to Alex for the challenging and interesting topic, for always guiding me in the right direction, and for patiently proofreading the whole text – carefully spotting every single little inaccuracy.

Thanks to my family for being a northern star, which I expect not to be as constant, but rather to shift with me across the constellations of life. My parents, my brother, my dog sister; my cousins, and all the relatives and friends which walk with me everyday in a gene or a gesture or a WhatsApp message.

Last but not least, many thanks to Matteo and Nika. Thanks Matteo for taking care of me and feeding me, specifically feeding me the italianity which I longed for when away from home.

And Nika, for making me question my beliefs, for growing up with me, for being the most breathtaking drug I've swallowed in my life.

* * *

Per prima cosa, la mia gratitudine va ad Alex, per la sfida che mi ha fornito con il tema di questa tesi, per guidarmi sempre nella direzione giusta, e per leggere con pazienza questo testo parola per parola – indicandomi con cura ogni imprecisione.

Grazie alla mia famiglia per essere una stella polare, che mi aspetto non altrettanto costante, ma che piuttosto scivoli con me attraverso le costellazioni della vita. I miei genitori, mio fratello, la mia sorella cane; le mie cugine, tutti i parenti e gli amici e le amiche che ogni giorno camminano con me in un gene o in gesto o in un messaggio WhatsApp.

Ultimi ma non per importanza, grazie mille a Matteo e Nika. Grazie Matteo per prendersi cura di me e darmi da mangiare, in particolare nutrirmi di quell'italianità di cui avevo fame quando lontano da casa.

E a Nika, per mettere in discussione quello in cui credo, per crescere insieme a me, per essere la droga più mozzafiato che io abbia mandato giù nella mia vita.

# Kurzfassung

Die Schnittelimination, eines der bekanntesten Probleme in der Beweistheorie, wurde für Sequenzenkalüle erster Ordnung von Gentzen in seinem gefeierten *Hauptsatz* definiert und gelöst.

Ceres bezeichnet einen anderen Algorithmus zur Schnittelimination erster und höherer Ordnung in der klassischen Logik. Er beruht auf der Idee einer charakteristischen Klauselmenge, die aus einem Beweis des Sequenzenkalküls extrahiert wurde und stets widerlegbar bleibt. Eine Resolutionswiderlegung dieser Klauselmenge dient als Gerüst für einen Beweis, der lediglich atomare Schnitte enthält. Das wird erreicht, indem die Klauseln der Resolutionswiderlegung durch die entsprechenden Beweisprojektionen des Originalbeweises ersetzt werden.

Ceres wurde auf Beweisschemata ausgedehnt, die als Schablonen für gewöhnliche Beweise erster Ordnung dienen, die natürliche Zahlen als Parameter haben. Jede Instantiierung der Parameter durch konkrete Zahlen ergibt einen neuen Beweis erster Ordnung. Das Anwenden existierender Algorithmen der Schnittelimination auf jeden Beweis dieser unendlichen Sequenz würde zu einer ebenfalls unendlichen Folge von schnittfrei Beweisen führen. Das Ziel in der Logik erster Ordnung von Ceres ist es nun, stattdessen eine einheitliche, schematische Beschreibung dieser Sequenz schnittfrei Beweise zu liefern. Um dieses Ziel zu erreichen, wurde jedes Konzept in Ceres schematisch entworfen: es enthält Schemata der charakteristischen Klauselmenge, Schemata der Resolutionswiderlegung, Projektionsschemata etc.

Während Ceres ein vollständiger Algorithmus zur Schnittelimination in der Logik erster Ordnung gilt, ist nicht klar, ob dies auch für die Schemata erster Ordnung zutrifft: liefert Ceres stets ein Schema schnittfreier Beweise, wenn ein Beweisschema mit Schnitten eingegeben wird? Die Schwierigkeit besteht darin, ein passendes Widerlegungsschema für das charakteristische Termschema eines Beweisschemas zu finden und darzustellen.

In der vorliegenden Arbeit beschäftigen wir uns mit der Lösung dieses Problems, indem wir Ceres auf aussagenlogische Schemata einschränken, welche als Schablonen für aussagenlogische Beweise dienen. Durch die adäquate Beschränkung der Aussagekraft der aussagenlogischen Schemata und Beweisschemata wollen wir eine Version schematischen Ceres' vorlegen, welche einen vollständigen Algorithmus der Schnittelimination für aussagenlogische Schemata liefert. Wir konzentrieren uns dabei auf einen bestimmten

Schritt von Ceres: Schemata der Resolutionswiderlegung. Zuerst beweisen wir, dass durch das einfache Adaptieren von Ceres für Schemata erster Ordnung für unseren Fall der Algorithmus unvolständig ist. Danach modifizieren wir das Konzept des Schemas der Resolutionswiderlegung: um eine Klauselmenge zu widerlegen, bringen wir sie zuerst in eine allgemeine Form, um im Anschluss eine festgelegte Widerlegung dieser allgemeinen Klauselmenge zu benutzen. Unsere Variation von schematischem Ceres stellt den ersten Schritt in Richtung einer Vollständigkeit aussagenlogischer Schemata dar.

# Abstract

Cut-elimination is one of the most famous problems in proof theory, and it was defined and solved for first-order sequent calculus by Gentzen in his celebrated *Hauptsatz*.

Ceres is a different cut-elimination algorithm for first- and higher-order classical logic. It is based on the notion of a characteristic set of clauses which is extracted from a proof in sequent calculus and is always unsatisfiable. A resolution refutation of this clause set is used as a skeleton for a proof with only atomic cuts. This is obtained by replacing clauses from the resolution refutation with the corresponding proof projection derived from the original proof.

Ceres was extended to proof schemata, which are templates for usual first-order proofs, with parameters for natural numbers. Every instantiation of the parameters to concrete numbers yields a new first-order proof. We could apply existing algorithms for cut-elimination to each proof in this infinite sequence, obtaining an infinite sequence of cut-free proofs. The goal of Ceres for first-order schemata is instead to give a uniform, schematic description of this sequence of cut-free proofs. To this aim, every concept in Ceres was made schematic: there are characteristic clause set schemata, resolution refutation schemata, projection schemata, etc.

However, while Ceres is known to be a complete cut-elimination algorithm for first-order logic, it is not clear whether this holds for first-order schemata too: given in input a proof schema with cuts, does Ceres always produce a schema for cut-free proofs? The difficult step is finding and representing an appropriate refutation schema for the characteristic term schema of a proof schema.

In this thesis, we progress in solving this problem by restricting Ceres to propositional schemata, which are templates for propositional proofs. By limiting adequately the expressivity of propositional schemata and proof schemata, we aim at providing a version of schematic Ceres which is a complete cut-elimination algorithm for propositional schemata. We focus on one particular step of Ceres: resolution refutation schemata. First, we prove that by naively adapting Ceres for first-order schemata to our case, we end up with an incomplete algorithm. Then, we modify slightly the concept of resolution refutation schema: to refute a clause set, first we bring it to a generic form, and then we use a fixed refutation of that generic clause set. Our variation of schematic Ceres is the first step towards completeness with respect to propositional schemata.

# Contents

# Introduction

The aim of this Master's Thesis is to apply the *Ceres* method to *propositional proof schemata.*

In order to understand the main concepts, let us step back and examine first what are propositional schemata; we then move to cut elimination for schemata and, finally, to the Ceres method. At the end of this introduction, we will outline the structure of the remaining thesis.

A **propositional schema** (Definition 38) is a "template" for constructing propositional formulas. For example, the schema

$$(P(n) \ \wedge \ P(n) \supset P(n+1)) \ \supset \ P(n+1)$$

depends on the parameter $n$, which is a *placeholder* for a natural number. For every assignment of a number to $n$, we obtain a new propositional formula:

$$\begin{array}{c} (P(0) \ \wedge \ P(0) \supset P(1)) \ \supset \ P(1) \\ (P(1) \ \wedge \ P(1) \supset P(2)) \ \supset \ P(2) \\ (P(2) \ \wedge \ P(2) \supset P(3)) \ \supset \ P(3) \\ \vdots \end{array}$$

Thus propositional schemata are able to represent compactly an infinite number of propositional formulas.

Propositional schemata have many fruitful applications: one is *circuit verification*, since one usually models circuit verification problems as sequences of propositional problems, parameterized by a natural number $n$ (which encodes the number of bits of the data) [GF93]. Propositional schemata are also relevant to specify *problems over finite domains* (where the parameter $n$ is now the size of the domain): for example, the *pigeonhole principle*, *coloring graph* problems, and the *n-queens* problem [MNR$^+$08].

The advantage of representing an infinite number of formulas in a condensed way, is that one does not need anymore to prove separately every instance of the problem: it is enough to give, once and for all, a *proof schema* (Definition 55), which again is a template for usual proofs, and denotes an infinite sequence of proofs.

In [DLRW13], proof schemata are indeed used to express infinite proof sequences, but with an additional reason: to avoid using explicitly the *induction principle*. Induction is the inference rule used in sequent calculus to implement mathematical induction:

$$\frac{F(x), \Gamma \vdash \Delta, F(x+1)}{F(0), \Gamma \vdash \Delta, F(t)}\, ind$$

This inference characterizes arithmetic, but it must be circumvented because it interacts negatively with the traditional algorithms for **cut-elimination**. We will discuss cut-elimination in Section 1.4.1; for now, it suffices to know that *cut* is one of the structural rules of Gentzen's sequent calculus, and it has the form:

$$\frac{\Gamma \vdash \Delta, F \qquad F, \Pi \vdash \Lambda}{\Gamma, \Pi \vdash \Delta, \Lambda}\, cut$$

One of the major achievements in proof theory is the so-called *cut-elimination theorem*, asserting that every first-order proof which contains *cut* inferences can be transformed into another proof with no occurrences of cuts. The possibility of removing all cuts is a very strong statement: cut-elimination for first-order logic entails consistency for first-order sequent calculus, and it has a crucial role in Gentzen's proof of the *consistency of arithmetic*, a very hard problem. Sadly, it is not possible to extend Gentzen's algorithm for cut-elimination to Peano arithmetic, as we will see in Section 2.1.

Since cut-elimination cannot work with induction, an attractive approach is to get rid of the induction rule, and introduce *proof schemata* instead. The resulting logical system does not loose expressivity, because induction can be simulated by proof schemata: while the induction inference explicitly deals with an infinite number of cases, proof schemata do so in an implicit way. Remember that a proof schema is a shorthand for an infinite sequence of proofs. For every proof in this sequence, one could apply usual cut-elimination, and obtain an infinite sequence of cut-free proofs. But our wish is again to specify this infinite sequence by finite means. Therefore, new schematic formalisms are necessary, in order to represent the operations and the output of eliminating cuts from schemata.

That's where *Ceres* comes into play (see Section 1.4). Ceres stands for **Cut-Elimination by RESolution**, and it is – as the name suggests – a cut-elimination algorithm. In short, it reduces cut-elimination for a proof $\varphi$ to a theorem proving problem: finding a refutation of the characteristic clause set $\mathrm{CL}(\varphi)$. Given a *resolution refutation* (Definition 25) of that characteristic clause set, an essentially cut-free proof is then constructed through a simple proof-theoretic transformation.

Ceres was found to be particularly suited to generalize to schemata, and in fact it was ported to first-order schemata in [DLRW13]; similarly as we said above for the propositional case, first-order schemata are templates for first-order propositions. But unluckily, due to the high expressivity of first-order schemata, it is still not clear whether

Ceres for first-order schemata is a *complete* (see Theorem 1) cut-elimination method or not.

The goal of this thesis is to study instead a restricted problem: Ceres not for the full first-order schemata, but for the weaker propositional schemata. The original aim of our work was to prove completeness of Ceres for propositional logic (see Theorem 1); unfortunately, this was not possible, and in fact we will prove in Chapter 4 that completeness does not hold for our version of Ceres. Our Ceres is incomplete because our notion of *resolution proof schema* (which is the schematic version of resolution derivations, see Definition 59) is not strong enough to refute schematic clause terms. For this reason, we decided to follow another approach (Chapter 5), by slightly modifying the notion of *refutation schema*. This method proved successful: it allows to mechanically refute every *characteristic term schema* (Definition 66), which is the principal step of the Ceres method. The remaining steps of the algorithm – schematic proof projections and final *atomic cut normal form* – still needs to be investigated in the future.

This thesis is structured as follows:

- In Chapter 1, we present the basic notions and definitions, in order to make this thesis as self-contained as possible: propositional logic, sequent calculus, resolution calculus, and the Ceres method;
- In Chapter 2, we introduce propositional schemata. We start with a discussion on cut-elimination in arithmetic, which gives the motivation for the use of schemata. We then adapt the definitions in Chapter 1 to support schemata;
- Chapter 3 is a warm up, and it provides two clause sets which will prove useful for the results to follow;
- Chapter 4 contains some negative results: we discuss the completeness of Ceres for propositional schemata with respect to different forms of *resolution proof schemata*;
- The most interesting result is in Chapter 5, where we provide a new notion of *refutation schema*, which partially solves our original goal;
- Chapter 6 consists of an example of proof schema with which we test the method in Chapter 5;
- We finally summarize the main results and highlight possible future work in Chapter 7.

# CHAPTER 1

# Preliminaries

In this chapter, we are going to give the fundamental definitions which will be essential in the next chapters. We will define *propositional logic*, *sequent calculus*, and a *resolution calculus*. In the end, we will outline what is the problem of *cut-elimination*, and what is the *Ceres* method.

## 1.1 Propositional Logic

We introduce propositional logic in a similar way as in [Tak13]. We slightly modify the syntax of propositional logic to handle a different form of propositional atoms, that will come useful when we define *propositional schemata* in Section 2.2. Our propositional atoms will be for example $P(0)$, $P(12)$ or $Q(7)$: they are so-called *indexed atoms*, which consists of a natural number or *index*, in addition to a propositional symbol.

### 1.1.1 Syntax

**Definition 1** (Language)**.** The language of classical propositional logic consists of the following elements:

- a countably infinite number of *propositional symbols*, which we will denote by $P$, $Q$, etc.;
- the logical symbols $\wedge$ (and), $\vee$ (or), $\neg$ (not) and $\supset$ (implies);
- the natural numbers $0, 1, 2, \ldots$, whose set is denoted by $\mathbb{N}$.

**Definition 2** (Indexed proposition)**.** An *indexed proposition* is an expression of the form $P(\alpha)$, where $P$ is a propositional symbol, and $\alpha$ is a natural number.

**Definition 3** (Propositional formula)**.** The set of *formulas* is inductively defined as follows:

- each indexed proposition is a formula;
- if $F$ is a formula, then $\neg F$ is a formula;
- if $F_1$ and $F_2$ are formulas, then $F_1 \vee F_2$, $F_1 \wedge F_2$, and $F_1 \supset F_2$ are formulas.

### 1.1.2 Semantics

Our different definition of propositional atoms necessarily alters the definition of *propositional interpretation*: an interpretation is not just a mapping from propositional symbols to truth values as in [Tak13], but it must now take into account the natural numbers which index propositional atoms.

**Definition 4** (Interpretation)**.** An *interpretation of the propositional language* $\mathcal{I}$ is a function mapping every propositional symbol $P$ to a function $\mathcal{I}_P \colon \mathbb{N} \to \{\text{true}, \text{false}\}$.

**Definition 5** (Truth value)**.** Let $F$, $F_1$, $F_2$ be propositional formulas. The *truth value* $\mathcal{I}(F)$ of a propositional formula $F$ in an interpretation $\mathcal{I}$ is inductively defined as:

- $\mathcal{I}(P(\alpha)) = \mathcal{I}_P(\alpha)$, where $P$ is a propositional symbol and $\alpha$ a natural number;
- $\mathcal{I}(\neg F) = \text{true}$ iff $\mathcal{I}(F) = \text{false}$;
- $\mathcal{I}(F_1 \wedge F_2) = \text{true}$ iff $\mathcal{I}(F_1) = \text{true}$ and $\mathcal{I}(F_2) = \text{true}$;
- $\mathcal{I}(F_1 \vee F_2) = \text{true}$ iff $\mathcal{I}(F_1) = \text{true}$ or $\mathcal{I}(F_2) = \text{true}$;
- $\mathcal{I}(F_1 \to F_2) = \text{true}$ iff $\mathcal{I}(F_1) = \text{false}$ or $\mathcal{I}(F_2) = \text{true}$.

**Definition 6** (Model)**.** Let $F$ be a formula, and $\mathcal{I}$ be a propositional interpretation. We call $\mathcal{I}$ a model of $F$ if and only if $\mathcal{I}(F) = \text{true}$.

We denote that $\mathcal{I}$ is a model of $F$ by $\mathcal{I} \models F$.

**Definition 7** (Satisfiability)**.** Let $F$ be a formula.

- $F$ is called *satisfiable* if $F$ has a model.
- $F$ is called *unsatisfiable* if it is not satisfiable.
- $F$ is called *valid* if every interpretation is a model of $F$.

## 1.2 Sequent Calculus

Let us now define our calculus for propositional logic: it is nothing more than the calculus **LK** for classical logic, restricted to the propositional case (no quantifier inferences).

Definitions 8 and 9 are common knowledge.

**Definition 8** (Multiset)**.** *Multisets*, unlike usual sets, allow multiple instances of their elements. We denote multisets with the delimiters $\wr$ and $\int$.

For example, $\wr a, a, b \int$ is the multiset containg $a$ with multiplicity 2, and $b$ with multiplicity 1.

**Definition 9** (List)**.** *Lists*, like multisets, allow multiple instances of their elements, but unlike multisets, distinguish the order of their elements.

*Remark*: we assume an intuitive understanding of the symbols $\in$, $\subseteq$, and $\cup$ for multisets and lists.

**Definition 10** (Sequent)**.** Let $\Gamma$ and $\Delta$ be finite lists of formulas. The expression $\Gamma \vdash \Delta$ is called a *sequent.* $\Gamma$ is called the antecedent of $S$, and $\Delta$ the consequent of $S$. "$\vdash$" is called the *empty sequent.*

*Remark*: if $\Gamma$ and $\Pi$ are lists of formulas, we denote by $\Gamma, \Pi$ the concatenation of the two lists.

**Definition 11** (Semantics of sequents)**.** Let $S = A_1, \dots, A_\alpha \vdash B_1, \dots, B_\beta$ be a sequent. Then the semantics of $S$ can be expressed by the propositional formula:

$$(A_1 \wedge \dots \wedge A_\alpha) \rightarrow (B_1 \vee \dots \vee B_\beta) .$$

**Definition 12** (Composition of Sequents)**.** Let $S = \Gamma \vdash \Delta$ and $S' = \Pi \vdash \Lambda$ be two sequents. We define the composition of $S$ and $S'$ by $S \circ S'$, where $S \circ S' = \Gamma, \Pi \vdash \Delta, \Lambda$.

The usual notion of subsequent, adapted to our definition of sequents with lists of formulas.

**Definition 13** (Subsequent)**.** Let $S = \Gamma \vdash \Delta$ and $S' = \Pi \vdash \Lambda$ be sequents. We define $S \sqsubseteq S'$ if and only if $\text{multiset}(\Gamma) \subseteq \text{multiset}(\Pi)$ and $\text{multiset}(\Delta) \subseteq \text{multiset}(\Lambda)$, and call $S'$ a *subsequent* of $S$.

*Remark*: if $\Gamma$ is a list, by multiset($\Gamma$) we mean the multiset obtained from $\Gamma$ by ignoring the order of the elements.

**Example 1.** Let

$$\begin{aligned} S_1 &:= P(0), Q(1) \vdash Q(2) \\ S_2 &:= P(0), Q(1), R(2) \vdash P(1) \wedge R(3), Q(2) \end{aligned}$$

be two sequents. The first sequent is a subsequent of the second one, because

$$\begin{aligned} \wr P(0), Q(1) \int \ &\subseteq \wr P(0), Q(1), R(2) \int, \text{ and} \\ \wr Q(2) \int \ &\subseteq \wr P(1) \wedge R(3), Q(2) \int . \end{aligned}$$

The notion of initial sequent is as in [Tak13], adapted to our notion of propositional atom.

**Definition 14** (Initial sequent)**.** An *initial sequent* (or *axiom*) is a sequent of the form $P(\alpha) \vdash P(\alpha)$, where $P(\alpha)$ is a propositional atom.

We define the rules of inference of sequent calculus, as in [Tak13].

**Definition 15** (Inference)**.** An *inference* is an expression of the form

$$\frac{S_1}{S} \qquad \text{or} \qquad \frac{S_1 \quad S_2}{S},$$

where $S_1$, $S_2$ and $S$ are sequents. $S_1$ and $S_2$ are called *upper sequents*, and $S$ is called the *lower sequent* of the inference. Inferences with a single premise are called *unary*, those with two premises are called *binary*.

We restrict ourselves to inferences obtained from the following rules of inference, in which $F$ and $G$ denote formulas, and $\Gamma$, $\Pi$, $\Delta$ and $\Lambda$ denote lists of formulas:

- Structural rules:
  - Weakening

$$\frac{\Gamma \vdash \Delta}{F, \Gamma \vdash \Delta}\, w\colon l \qquad \frac{\Gamma \vdash \Delta}{\Gamma \vdash \Delta, F}\, w\colon r$$

    $F$ is called the *weakening formula*.
  - Contraction

$$\frac{F, F, \Gamma \vdash \Delta}{F, \Gamma \vdash \Delta}\, c\colon l \qquad \frac{\Gamma \vdash \Delta, F, F}{\Gamma \vdash \Delta, F}\, c\colon r$$

  - Exchange

$$\frac{\Gamma, F, G, \Pi \vdash \Delta}{\Gamma, G, F, \Pi \vdash \Delta}\, x\colon l \qquad \frac{\Gamma \vdash \Delta, F, G, \Lambda}{\Gamma \vdash \Delta, G, F, \Lambda}\, x\colon r$$

  Weakening, contraction and exchange are called *weak inferences*, while all others will be called *strong inferences*.

– Cut

$$\frac{\Gamma \vdash \Delta, F \qquad F, \Pi \vdash \Lambda}{\Gamma, \Pi \vdash \Delta, \Lambda}\,cut$$

$F$ is called the *cut formula* of this inference.

- Logical rules:

– $\neg$ (negation)

$$\frac{\Gamma \vdash \Delta, F}{\neg F, \Gamma \vdash \Delta}\,\neg\colon l \qquad \frac{F, \Gamma \vdash \Delta}{\Gamma \vdash \Delta, \neg F}\,\neg\colon r$$

$F$ and $\neg F$ are called the *auxiliary formula* and the *principal formula*, respectively, of this inference.

– $\wedge$ (conjunction)

$$\frac{F, \Gamma \vdash \Delta}{F \wedge G, \Gamma \vdash \Delta}\,\wedge\colon l_1 \quad \text{and} \quad \frac{G, \Gamma \vdash \Delta}{F \wedge G, \Gamma \vdash \Delta}\,\wedge\colon l_2$$

$$\frac{\Gamma \vdash \Delta, F \qquad \Gamma \vdash \Delta, G}{\Gamma \vdash \Delta, F \wedge G}\,\wedge\colon r$$

$F$ and $G$ are called the *auxiliary formulas* and $F \wedge G$ is called the *principal formula* of this inference.

– $\vee$ (disjunction)

$$\frac{F, \Gamma \vdash \Delta \qquad G, \Gamma \vdash \Delta}{F \vee G, \Gamma \vdash \Delta}\,\vee\colon l$$

$$\frac{\Gamma \vdash \Delta, F}{\Gamma \vdash \Delta, F \vee G}\,\vee\colon r_1 \quad \text{and} \quad \frac{\Gamma \vdash \Delta, G}{\Gamma \vdash \Delta, F \vee G}\,\vee\colon r_2$$

$F$ and $G$ are called the *auxiliary formulas* and $F \vee G$ is called the *principal formula* of this inference.

– $\supset$ (implication)

$$\frac{\Gamma \vdash \Delta, F \qquad G, \Pi \vdash \Lambda}{F \supset G, \Gamma, \Pi \vdash \Delta, \Lambda}\,\supset\colon l \quad \text{and} \quad \frac{F, \Gamma \vdash \Delta, G}{\Gamma \vdash \Delta, F \supset G}\,\supset\colon r$$

$F$ and $G$ are called the *auxiliary formulas* and $F \supset G$ is called the *principal formula* of this inference.

We now define *propositional proofs*, which are LK-proofs of Definition 2.2 in [Tak13].

**Definition 16** (Proof)**.** We denote a proof $\varphi$ with end-sequent $S$ by:

$$\begin{array}{c}\varphi\\ \vdots\\ S\end{array} \quad \text{or} \quad \begin{array}{c}(\varphi)\\ S\end{array} \quad \text{or simply} \quad \varphi \ .$$

The set of proofs is inductively defined as follows:

- an initial sequent $S$ is a proof with end-sequent $S$;
- if $\varphi'$ is a proof with end-sequent $S'$, and $\frac{S'}{S}$ is a unary inference, then

$$\begin{array}{c}(\varphi')\\ \dfrac{S'}{S}\end{array}$$

  is a proof with end-sequent $S$;
- if $\varphi_1$ is a proof with end-sequent $S_1$, $\varphi_2$ is a proof with end-sequent $S_2$, and $\frac{S_1 \quad S_2}{S}$ is a binary inference, then

$$\dfrac{\begin{array}{c}(\varphi_1)\\ S_1\end{array} \qquad \begin{array}{c}(\varphi_2)\\ S_2\end{array}}{S}$$

  is a proof with end-sequent $S$.

**Definition 17** (Cut ancestor)**.** Auxiliary formulas are *immediate ancestors* of their principal formula. *Ancestors* are defined as the reflexive and transitive closure of the relation of immediate ancestor.

The occurrence of a formula in a proof is called a *cut ancestor* if it is the ancestor of a cut formula.

We now define *configurations*, which are objects keeping track of formulas in sequents. Configurations designate subsequents of a given sequent, and they will be used in the Ceres method to keep track of formulas which are cut-ancestors in a proof.

**Definition 18** (Configuration)**.** Let $S = F_1, \ldots, F_\alpha \vdash G_1, \ldots, G_\beta$ be a sequent. A configuration $\Omega$ of $S$ is an expression of the form $\diamond, \ldots, \diamond \vdash \diamond, \ldots, \diamond$, where:

- there are $\alpha$ diamonds on the left of $\vdash$,
- there are $\beta$ diamonds on the right of $\vdash$, and

- each diamond $\diamond$ can be either a $\square$ or a $\blacksquare$.

By $S \cdot \Omega$ we denote the subsequent of $S$ obtained from $S$ by keeping the formulas whose position in $S$ corresponds in $\Omega$ to a "$\blacksquare$", and removing the formulas whose position in $S$ corresponds in $\Omega$ to a "$\square$".

Intuitively, $\blacksquare$ means that the formula is selected, and $\square$ means that the formula is ignored.

**Example 2.** Let $S = P(0), Q(1) \vdash R(2)$; then we have:

- $S \cdot (\square, \square \vdash \square) = \ \vdash$;
- $S \cdot (\square, \blacksquare \vdash \square) = Q(1) \vdash$;
- $S \cdot (\blacksquare, \square \vdash \blacksquare) = P(0) \vdash R(2)$.

We need a way of keeping track of the formula positions in a configuration, when we modify the relative sequent by inferences. This is why we introduce the notion of *induced configuration*:

**Definition 19** (Induced configuration)**.** Let $S$, $S'$, $S_1$ and $S_2$ be sequents, $\Omega$ a configuration for $S$, and $\xi$ an inference.

We define the *configuration induced by* $\Omega$ on $S'$ (resp. $S_1$ and $S_2$) according to $\xi$:

- $\xi$ is a unary inference
  - If $\xi$ is $w\colon l$

    $$\frac{\Gamma \vdash \Delta}{F, \Gamma \vdash \Delta}\, w\colon l$$

    then the configuration $\diamond_0, \diamond_1, \ldots, \diamond_\alpha \vdash \diamond'_1, \ldots, \diamond'_\beta$ on $F, \Gamma \vdash \Delta$ induces the configuration $\diamond_1, \ldots, \diamond_\alpha \vdash \diamond'_1, \ldots, \diamond'_\beta$ on $\Gamma \vdash \Delta$ (and vice versa);
    Similarly for $w\colon r$;
  - If $\xi$ is $c\colon l$

    $$\frac{F, F, \Gamma \vdash \Delta}{F, \Gamma \vdash \Delta}\, c\colon l$$

    then the configuration $\diamond_0, \diamond_1, \ldots, \diamond_\alpha \vdash \diamond'_1, \ldots, \diamond'_\beta$ on $F, \Gamma \vdash \Delta$ induces the configuration $\diamond_0, \diamond_0, \diamond_1, \ldots, \diamond_\alpha \vdash \diamond'_1, \ldots, \diamond'_\beta$ on $F, F, \Gamma \vdash \Delta$ (and vice versa);
  - Similarly for the other unary inferences: $c\colon r$, $x\colon l$, $x\colon r$, $\neg\colon l$, $\neg\colon r$, $\wedge\colon r$, $\vee\colon l$, and $\supset\colon r$.
- If $\xi$ is a binary inference

– If $\xi$ is a *cut*:

$$\frac{\Gamma \vdash \Delta, F \qquad F, \Pi \vdash \Lambda}{\Gamma, \Pi \vdash \Delta, \Lambda}\, cut$$

then the configuration $\diamond_1, \ldots, \diamond_{\alpha_1}, \diamond'_1, \ldots, \diamond'_{\alpha_2} \vdash \diamond''_1, \ldots, \diamond''_{\beta_1}, \diamond'''_1, \ldots, \diamond'''_{\beta_2}$ on $\Gamma, \Pi \vdash \Delta, \Lambda$ induces the configuration $\diamond_1, \ldots, \diamond_{\alpha_1} \vdash \diamond''_1, \ldots, \diamond''_{\beta_1}, \blacksquare$ on $\Gamma \vdash \Delta, F$ and $\blacksquare, \diamond'_1, \ldots, \diamond'_{\alpha_2} \vdash \diamond'''_1, \ldots, \diamond'''_{\beta_2}$ on $F, \Pi \vdash \Lambda$ (and vice versa);

– Similarly for the other binary inferences: $\wedge\colon l$, $\vee\colon r$, and $\supset\colon l$.

**Example 3.** Suppose we have the following inference:

$$\frac{A, B, C \vdash D}{A, B \vdash C \supset D}\supset\colon r$$

The configuration $\blacksquare, \square \vdash \blacksquare$ on $A, B \vdash C \supset D$ induces the configuration $\blacksquare, \square, \blacksquare \vdash \blacksquare$ on $A, B, C \vdash D$ (and vice versa).

In the usual version of Ceres for first-order logic, configurations are denoted directly on sequents. For example:

$$\frac{A^\star, B, C^\star \vdash D^\star}{A^\star, B \vdash (C \supset D)^\star}\supset\colon r$$

In this thesis, instead, we prefer to give a more precise treatment of configurations.

## 1.3 Resolution Calculus

Our formulation of the resolution calculus operates on specific sequents – called clauses – and uses the only rule of *resolution.* A second version of the resolution calculus with an additional rule of *weakening* is introduced at the end of the section.

For the definitions in this section, we adapt the notion of $p$-resolution deduction in [BL06] and [BL11].

**Definition 20** (Clause)**.** A sequent is called *atomic* or a *clause* if it contains only atomic formulas.

**Definition 21** (Tautology)**.** A clause $C$ is called a *tautology* if there is a propositional atom which occurs both in the antecedent and in the consequent of $C$.

The definition of *subsumption* is adapted from [BL11], taking into account our definition of sequents with lists.

**Definition 22** (Subsumption)**.** A clause $C = \Gamma \vdash \Delta$ subsumes a clause $D = \Pi \vdash \Lambda$ ($C \leq D$) if and only if $\mathrm{set}(\Gamma) \subseteq \mathrm{set}(\Pi)$ and $\mathrm{set}(\Delta) \subseteq \mathrm{set}(\Lambda)$.

Note: we define $\mathrm{set}(\Gamma) := \{A \mid A \in \Gamma\}$.

**Definition 23** (Resolvent)**.** Let $C = \Gamma \vdash \Delta$ and $D = \Pi \vdash \Lambda$ be clauses, and $A$ be a propositional atom. Then the *resolvent* of $C$ and $D$ with respect to $A$ is

$$\mathrm{res}(C, D; A) = \Gamma, \Pi' \vdash \Delta', \Lambda,$$

where $\Pi'$ and $\Delta'$ are respectively $\Pi$ and $\Delta$ without all the occurrences of the atom $A$.

*Remark*: from the definition of resolvent, it is clear that we avoid the need of contractions in the resolution calculus, by actually treating clauses as *sets* of atoms.

**Definition 24** (Resolution deduction)**.** Let $\mathcal{C}$ be a set of clauses. We define inductively the set of *resolution deductions* by:

- a clause $C$ in $\mathcal{C}$ is a resolution deduction from $\mathcal{C}$ with end-clause $C$;
- if $\gamma_1$ and $\gamma_2$ are resolution deductions from $\mathcal{C}$ with end-clauses respectively $C_1$ and $C_2$, and $\mathrm{res}(C_1, C_2, A) = C$, then $\mathrm{r}(\gamma_1, \gamma_2; A)$ is a resolution deduction from $\mathcal{C}$ with end-clause $C$.

**Definition 25** (Resolution refutation)**.** A resolution deduction from $\mathcal{C}$ with end-clause $\vdash$ is called a *resolution refutation*.

**Example 4.** Let us consider the following resolution tree:

$$\dfrac{\vdash A \qquad \dfrac{A \vdash B \qquad B \vdash}{A \vdash}}{\vdash}$$

This tree can be formalized in the following resolution deduction:

$$\mathrm{r}((\vdash A), \mathrm{r}((A \vdash B), (B \vdash); B); A).$$

### 1.3.1 Resolution with weakenings

In this paragraph, we define a slight variation of the resolution calculus which we defined in Section 1.3. We call this new calculus *w-resolution* because it is obtained by allowing also *weakening* steps in resolution deductions (besides usual resolution steps).

**Definition 26** ($w$-resolution deduction)**.** Let $\mathcal{C}$ be a set of clauses. We define inductively the set of $w$-resolution deductions similarly as in Definition 24:

- a clause $C$ in $\mathcal{C}$ is a $w$-resolution deduction from $\mathcal{C}$ with end-clause $C$;

- if $\gamma_1$ and $\gamma_2$ are $w$-resolution deductions from $\mathcal{C}$ with end-clauses respectively $C_1$ and $C_2$, and $\mathrm{res}(C_1, C_2, A) = C$, then $\mathrm{r}(\gamma_1, \gamma_2; A)$ is a $w$-resolution deduction from $\mathcal{C}$ with end-clause $C$.
- if $\gamma$ is a $w$-resolution deduction from $\mathcal{C}$ with end-clause $C$, and $D$ is any clause, then $w(\gamma; D)$ is a $w$-resolution term with end-clause $C \circ D$.

**Definition 27** (Weakening term)**.** $w$-resolution terms of the form $w(\gamma; D)$ are called *weakening terms.*

**Example 5.** Similarly as in the previous example, let us consider the following resolution tree where we allow a *weakening inference* $w$:

$$\dfrac{\vdash A \qquad \dfrac{\dfrac{A \vdash}{A \vdash B}\,w \qquad B \vdash}{A \vdash}}{\vdash}$$

The tree above can now be formalized in the following resolution deduction:

$$\mathrm{r}((\vdash A), \mathrm{r}(w(A \vdash; \vdash B), (B \vdash); B); A).$$

## 1.4 Cut-Elimination and Ceres

### 1.4.1 The Problem of Cut-Elimination

As we already saw in Definition 15, *cut* is the following rule of inference:

$$\dfrac{\Gamma \vdash \Delta, F \qquad F, \Pi \vdash \Lambda}{\Gamma, \Pi \vdash \Delta, \Lambda}\,cut$$

Gerhard Gentzen introduced the problem of reductive cut-elimination in his seminal papers [Gen35a] and [Gen35b]. Gentzen called his cut-elimination theorem "*Hauptsatz*", which means *main theorem*; and indeed his algorithm is a fundamental result in proof theory. Cut-elimination means that the *cut* rule can be removed from first-order sequent calculus – i.e. every proof with cuts can be transformed into another proof with the same end-sequent, but with no *cut* inferences.

Gentzen's procedure for cut elimination works by removing one cut at a time, and it is carried out by double induction on the *grade* and on the *rank* of the proof: the grade is the complexity of the cut formula, and the rank is the maximum number of consecutive sequents (counted upward) which contain the cut formula. A cut is eliminated either directly, by reducing the grade, or indirectly, by reducing the rank. One starts with the subproof containing only the uppermost *cut* inference, and proceeds stepwise until all the cuts are removed. Eliminating a cut is said to reduce the *cut complexity* of the proof, thus Gentzen's procedure is among the *reductive* methods.

Why is restricting the use of the cut rule so critical? After all, cuts may enable to make proof shorter. In fact, there are first-order logic proofs (with cuts) of length $n$, whose equivalent cut-free proofs have length

$$2^{2^{\cdot^{\cdot^{2}}}} \Big\} n \text{ times.}$$

This difference in length is enormous, and it comes without saying that for proofs "the smaller the better", both for a human to read them, and for computers to represent and check them [Ore82] [Pud98] [Sta79].

But actually there are different reasons to forbid the *cut* rule. The most important one concerns *automated theorem proving*: when searching for the proof of a sequent, one would hope to depend only on formulas obtained by breaking up syntactically the goal formulas. The *cut* inference, instead, requires one to pick up an appropriate formula (the cut formula) which is not necessarily related to the goal sequent. This enlarges dramatically the *search space*, making the automatic search for proofs harder.

This property of a logical system – the fact that only sub-formulas of the end-sequents are necessary for their proofs – is called *sub-formula property*, and systems with such a property are said to have *analytic proofs*.

Logical calculi which enjoy cut-elimination usually have the sub-formula property, and it is easy to see that this property actually implies the *consistency* of the calculus: in fact, suppose the empty sequent $\vdash$ could be proven. Then, in a proof of $\vdash$ would only occur sub-formulas of $\vdash$. But there are no sub-formulas of $\vdash$. Thus $\vdash$ cannot be derived – unless it is itself an axiom, which would be silly. Therefore, another valuable consequence of cut-elimination is that it easily proves consistency of the calculus. While consistency of first-order logic is not very tempting, Gentzen's method of cut-elimination was born to prove the consistency of *Peano arithmetic*, which is axiomatized in predicate logic. We will discuss cut-elimination for Peano arithmetic in Section 2.1.

The goal of proving consistency by cut-elimination was a very important problem in the history of proof theory, but today cut-elimination mainly benefits other areas, for example *proof analysis* (that consists in analyzing existing mathematical arguments) and *proof mining* (which extracts from proofs hidden mathematical information) [Lei15]. In fact, cut-elimination corresponds to removing lemmas from proofs, resulting in new direct proofs giving novel insights on theorems. A paradigmatic example of applying cut-elimination to concrete proofs is Girard's analysis in [Gir87] of Fürstenberg and Weiss' topological proof [FW78] of van der Waerden's theorem [van27] on partitions. The result of the analysis is astonishing: van der Waerden's original elementary proof can be obtained by applying cut-elimination to the proof of Fürstenberg and Weiss.

In the next section, we are going to introduce a different procedure for cut-elimination in first-order logic, called *Ceres*.

### 1.4.2 Cut-Elimination by Resolution

In the previous paragraph, we explained why and how to eliminate cuts from proofs; we also mentioned the most famous algorithm to accomplish cut-elimination, the one used by Gentzen to prove his *Hauptsatz*.

In [BL00], Matthias Baaz and Alexander Leitsch defined a totally different method of cut-elimination for first-order logic. As we noted above, Gentzen's reductive algorithm operates on proofs by removing one cut at a time. The method we outline in this section, instead, analyses *all* cut inferences in a proof at the same time, thus exhibiting better computational behaviour due to its global action [BL11].

Even though in this thesis we are going to restrict it to propositional logic, Ceres was originally formulated for classical first-order logic in [BL00] and [BL06]; later it was extended to higher-order logic in [HLW11]. Contrary to Gentzen's method, Ceres requires the original proof to be *Skolemized*: this means that *strong quantifiers* need to be removed in an initial phase, by replacing them with function symbols. In the Skolemized proof, the interplay of inferences which operate on ancestors of cut-formulas and of those which do not, produces a structure which can be represented as a set of clauses, or as a clause set term. The set of clauses extracted from a proof is always unsatisfiable, and therefore there exists a resolution refutation of it. This refutation (or, in the case of first-order logic, a ground instance of the refutation) is used as a skeleton of a proof with the same end-sequent as the original one, in which cuts are only on atomic formulas. The proof with atomic cuts is obtained by replacing the clauses in the resolution derivation with *proof projections* of the original proof.

The method Ceres – adapted to propositional logic – hence consists of the following steps:

1. construction of the characteristic clause set $\mathrm{CL}(\varphi)$;
2. computation of projections $\varphi(C)$ for each $C$ in $\mathrm{CL}(\varphi)$;
3. construction of a resolution refutation $\gamma$ of $\mathrm{CL}(\varphi)$;
4. merging the projections $\varphi(C)$ and the resolution refutation $\gamma$.

#### Clause Set Terms

In order to define the *characteristic clause set* of a propositional proof (the *Step 1* in the list above), we first introduce *clause set terms*, which represent clause sets preserving the structure of the originating proof.

Definitions 28 and 29 are originally from [BL11], but modified as in [DLRW13].

**Definition 28** (Clause set term)**.** *Clause set terms* are defined inductively by:

- $[C]$ is a clause set term, where $C$ is a clause;

- $X \oplus Y$ and $X \otimes Y$ are clause set terms, if $X$ and $Y$ are clause set terms.

**Definition 29** (Semantics of clause set terms)**.** We define a mapping $|\cdot|$ from clause set terms to sets of clauses in the following way:

- $|[C]| := \{C\}$, where $C$ is a clause,
- $|X \oplus Y| := |X| \cup |Y|$,
- $|X \otimes Y| := |X| \times |Y|$,

where $\mathcal{C} \times \mathcal{D} := \{C \circ D \mid C \in \mathcal{C}, D \in \mathcal{D}\}$.

The following is adapted from [BL11].

**Definition 30** (Characteristic term)**.** Let $\varphi$ be a propositional proof with end-sequent $S$, and $\Omega$ a configuration of $S$ (Definition 18).

We define the *characteristic (clause set) term* $\Theta^{\Omega}(\varphi)$ by induction on the structure of $\varphi$:

- If $\varphi$ consists of the initial sequent $S$, then $\Theta^{\Omega}(\varphi) := [S \cdot \Omega]$;
- If $\varphi$ ends with a unary inference:

$$\cfrac{\begin{matrix}(\varphi')\\ S'\end{matrix}}{S}$$

  then $\Theta^{\Omega}(\varphi) := \Theta^{\Omega'}(\varphi')$ where $\Omega'$ is the configuration induced by $\Omega$ on $S'$.

- If $\varphi$ ends with a binary inference:

$$\cfrac{\begin{matrix}(\varphi_1)\\ S_1\end{matrix} \qquad \begin{matrix}(\varphi_2)\\ S_2\end{matrix}}{S}$$

  - if the principal formula of the inference is in $\Omega$, or if the inference is a cut, then $\Theta^{\Omega}(\varphi) := \Theta^{\Omega_1}(\varphi_1) \oplus \Theta^{\Omega_2}(\varphi_2)$, where $\Omega_1$ and $\Omega_2$ are the configurations induced by $\Omega$ respectively on $S_1$ and $S_2$;
  - otherwise $\Theta^{\Omega}(\varphi) := \Theta^{\Omega_1}(\varphi_1) \otimes \Theta^{\Omega_2}(\varphi_2)$, where $\Omega_1$ and $\Omega_2$ are defined as above.

*Remark*: by $\Theta(\varphi)$ we mean $\Theta^{\varnothing}(\varphi)$, where $\varnothing$ is the empty configuration:

$$\varnothing := \Box, \ldots, \Box \vdash \Box, \ldots, \Box.$$

**Definition 31** (Characteristic clause set)**.** Let $\varphi$ be a propositional proof, and $\Theta(\varphi)$ be the characteristic term of $\varphi$. Then $\mathrm{CL}(\varphi) := |\Theta(\varphi)|$ is called the *characteristic clause set* of $\varphi$.

**Example 6.** See [BL11], Example 6.4.1.

The intuitive justification for Proposition 1 is the following. Every proof $\varphi$ can be transformed into a corresponding proof $\varphi'$ with empty end-sequent, by skipping the inferences which go into the end-sequent. $\mathrm{CL}(\varphi)$ contains the axioms used in the refutation $\varphi'$. For a complete proof, see [BL11], Proposition 6.4.1.

**Proposition 1.** $\mathrm{CL}(\varphi)$ is unsatisfiable.

### Proof Projections

Let us now turn to *Step 2*: we define *proof projections*, similarly as in [BL11]. The work on this thesis focuses on the refutation of characteristic clause sets of proofs, thus proof projections are not used and will not be defined in depth.

**Definition 32** (Projection)**.** Let $\varphi$ be a propositional proof, and $C \in \mathrm{CL}(\varphi)$. We define $\varphi(C)$, called the *projection* of $\varphi$ with respect to $C$.

We just give the intuitive idea; for the complete definition, please see [BL11], Lemma 6.4.1.

Proof projections are defined by replaying the original proof, and performing only the inferences on non-ancestors of cut formulas. In the case of a binary inference whose auxiliary formulas are ancestor of a cut formula, then one has to apply weakening in order to obtain the required formulas from the other premise [BL06].

**Example 7.** For an example of computation of the proof projections, see [BL11], Example 6.4.3.

Let us now define Ceres by combining what we introduced above. We know from Proposition 1 that every characteristic clause set $\mathrm{CL}(\varphi)$ is unsatisfiable. By the completeness of the resolution principle, there exists a resolution refutation $\gamma$ of $\mathrm{CL}(\varphi)$. By Definition 32, for every clause $C$ in the characteristic clause set, it is possible to build the corresponding proof projection $\varphi(C)$ with end-sequent $S \circ C$. Finally, just replace every occurrence of a clause $C$ in a leaf of the refutation $\gamma$ by $\varphi(C)$. The resulting proof has only atomic cuts, and end-sequent $S$.

It follows that:

**Theorem 1** (Completeness of Ceres for propositional logic)**.** Ceres is a cut-elimination method, i.e. for every propositional proof with end-sequent $S$, Ceres produces a corresponding propositional proof with end-sequent $S$ with only atomic cuts.

*Proof.* In [BL11], Theorem 6.4.1. See also the explanation above. □

To conclude, in the section above we defined Ceres for propositional logic, by adapting it from the original Ceres for first-order schemata in [DLRW13]. The main topic of this thesis is not just propositional proofs (which would be trivial), but rather infinite sequences of propositional proofs represented by *proof schemata*, as we will see in Section 2.2. Therefore, we are going to extend Ceres in Section 2.5 to propositional schemata, following [DLRW13].

CHAPTER 2

# Schemata

In this chapter, we are going to introduce the concept of *schemata*. In logic, schemata are usually expressions which use meta-variables that can be replaced by objects to yield well-formed formulas.

In our case, we use the metavariable $n$, which is a parameter standing for a natural number. For every concrete natural number $\alpha$, we may evaluate $n$ to $\alpha$ and obtain a ground object. Thus schemata allow us to describe an infinite sequence of objects (may they be formulas, proofs, resolution derivations, ...) by a finite description.

The starting point are *propositional schemata*, a simple schematic generalization of propositional logic (Section 2.2). As we are going to discuss in Section 2.1, propositional schemata and proof schemata are introduced to study the problem of cut-elimination in arithmetic from a different angle.

In the next sections, after a short introduction on cut-elimination in the presence of induction, we will provide the schematic variants of propositional logic, sequent calculus, resolution calculus, and Ceres.

## 2.1 Cut-elimination with Induction

In Section 1.4.1, we explored the problem of cut-elimination in first-order logic. As we said, the possibility of removing the cut rule from a calculus is rather important, because it has many fruitful applications to different areas of proof theory, like proof analysis, proof mining, etc.

Not all proofs can be formalized in first-order sequent calculus, and in fact many proofs in Mathematics need *induction*, which makes it possible to prove statements for *all* natural numbers, but in a finite way. Induction is one of the axioms which characterize arithmetic, and the theory of arithmetic was formalized by Gentzen in the system PA [Tak13]. PA stands for *Peano Arithmetic*, and it can be formulated in first-order sequent calculus:

- by adding extra initial sequents (called *mathematical initial sequents*): these axioms model the definition of *addition* and *multiplication*, and properties of the *successor*;
- by adding a new rule of inference, called *ind*:

$$\frac{F(x), \Gamma \vdash \Delta, F(x+1)}{F(0), \Gamma \vdash \Delta, F(t)}\ \mathit{ind}$$

  where $x$ is not in $F(0)$, $\Gamma$ or $\Delta$.

Unfortunately, reductive cut-elimination like Gentzen's algorithm is impossible in the presence of an induction rule like the one above [Tak13].

For example, in

$$\frac{\dfrac{F(x), \Gamma \vdash \Delta, F(x+1)}{F(0), \Gamma \vdash \Delta, F(t)}\ \mathit{ind} \qquad \begin{array}{c}\vdots\\ F(t), \Pi \vdash \Lambda\end{array}}{F(0), \Gamma, \Pi \vdash \Delta, \Lambda}\ \mathit{cut}$$

if $t$ is *not* a ground term, the *cut* cannot be shifted over the induction inference and thus cannot be eliminated (if instead $t$ is a ground term in the language of arithmetic, then it can be evaluated, and the *ind* removed [Tak13]).

See [Tai68] and [MM00] for other approaches to inductive reasoning using induction rules. Instead, we follow the approach in [DLRW13], and we replace the explicit *ind* rule, with *proof schemata*. In fact, *ind* inferences can be replaced by iterated cuts in proof schemata in the following (informal) way: suppose we have a proof schema $\varphi(n)$ with end-sequent $F(n), \Gamma \vdash \Delta, F(n+1)$; we want to simulate the *ind* rule, and give a proof schema with end-sequent $F(0), \Gamma \vdash \Delta, F(t)$. We define the proof schema $\boldsymbol{\psi}(n)$ with end-sequent $F(0), \Gamma \vdash \Delta, F(n)$ by *primitive recursion*, which means that the proof schema for $n+1$ can "depend" on the case for $n$. The base case $n = 0$ is just $\varphi(0)$, thus $\boldsymbol{\psi}(0) := \varphi(0)$. As for the inductive case, suppose we already have a proof $\boldsymbol{\psi}(n)$ with end-sequent $F(0), \Gamma \vdash \Delta, F(n)$; then $\boldsymbol{\psi}(n+1)$ is:

$$\dfrac{\dfrac{\dfrac{\begin{array}{c}\boldsymbol{\psi}(n)\\ \hdashline F(0), \Gamma \vdash \Delta, F(n)\end{array} \qquad \begin{array}{c}(\boldsymbol{\varphi}(n))\\ F(n), \Gamma \vdash \Delta, F(n+1)\end{array}}{F(0), \Gamma, \Gamma \vdash \Delta, \Delta, F(n+1)}\ \mathit{cut}}{\text{contractions}}}{F(0), \Gamma \vdash \Delta, F(n+1)}$$

As we can see above, $\boldsymbol{\psi}(n+1)$ depends recursively on $\boldsymbol{\psi}(n)$. By using a *cut* with $\varphi(n)$ on $F(n)$, plus many contractions, we prove $F(0), \Gamma \vdash \Delta, F(n+1)$. In conclusion, the required proof schema for $F(0), \Gamma \vdash \Delta, F(t)$ is simply $\boldsymbol{\psi}(t)$.

Now, since proof schemata are just a compact representation of infinite sequences of proofs, given a proof schema $\varphi$, we can compute for every natural number $\alpha$ its corresponding propositional or first-order proof $\varphi \downarrow \alpha$, and then apply a usual cut-elimination algorithm to it. But our goal is different: we'd rather want to describe the cut-free proofs for every $\alpha$ in a uniform way, with a parameter $n$.

One could actually try to naively apply to proof schemata usual reductive methods, like Gentzen-style cut-elimination, but this would not work, since it is not clear how to move the cut through a proof link:

$$\dfrac{\dfrac{\psi(n)}{\Gamma \vdash \Delta, F} \qquad \dfrac{\vdots}{F, \Pi \vdash \Lambda}}{\Gamma, \Pi \vdash \Delta, \Lambda}\ cut$$

This problem is not peculiar to our schematic calculus, but a general one for this kind of proofs (see [Bro05]).

This is why in this thesis we follow [DLRW13], and use the Ceres method instead. Ceres is particularly suited for generalizing to proof schemata, because all the steps of that algorithm can be naturally extended to the schematic case. As we will see in Section 2.5, porting Ceres to proof schemata would essentially solve the problem of cut-elimination in the presence of induction, since the output of the algorithm would be the schema for a proof with only atomic cuts.

## 2.2 Propositional Schemata

In this section, we introduce our calculus for propositional schemata. As we already noted, propositional schemata represent infinite sequences of usual propositional formulas, indexed by parameters which stand for natural numbers.

Our calculus is similar to the one introduced by Aravantinos and others in [ACP09]. In [ACP11], they proved that satisfiability for their version of propositional schemata is undecidable. In this thesis, we will use a modified version of those schemata. We restrict ourselves to *monadic* schemata too, i.e. the propositional atoms are indexed by one arithmetic expression; we use *linear arithmetic expressions* of the form $\alpha \times n + \beta$, permitting only one arithmetic variable, $n$.

Furthermore, while [ACP09] features only *iterated conjunctions and disjunctions* of the form $\bigwedge_{i=0}^{n} F$ and $\bigvee_{i=0}^{n} F$, our calculus allows *definitions*, i.e. the possibility of defining predicates by primitive recursive specifications (we will detail this in Definition 38). This increases the expressivity of the propositional calculus, and allows to represent more complex sequences of propositional formulas.

### 2.2.1 Syntax

We define the language of propositional schemata, similarly as in Definition 1.

**Definition 33.** The *language of propositional schemata* consists of the following elements:

- a countably infinite number of propositional symbols;
- the logical symbols $\wedge$ (and), $\vee$ (or), $\neg$ (not) and $\supset$ (implies);
- the natural numbers 0, 1, 2, ...
- the arithmetic variable $n$,
- the symbols "+" and "$\times$".

We choose to limit the expressivity of the calculus by indexing atomic formulas only with *linear* arithmetic expressions:

**Definition 34** (Arithmetic expression)**.** An *arithmetic expression* is a formal expression of the form $\alpha \times n + \beta$, where $\alpha$ and $\beta$ are natural numbers.

**Example 8.** Examples of arithmetic expressions are:

- $4 \times n + 3$, which we also write $4n + 3$,
- $0 \times n + 7$, which we also abbreviate with just 7.

**Definition 35** (Replacement in arithmetic expressions)**.** Let $a = \alpha_1 \times n + \beta_1$ and $b = \alpha_2 \times n + \beta_2$ be two arithmetic expressions. We denote by $a[n \mapsto b]$ the expression obtained by replacing $n$ with $b$:

$$a[n \mapsto b] := (\alpha_1 \cdot \alpha_2) \times n + (\alpha_1 \cdot \beta_2 + \beta_1).$$

Arithmetic expressions can be considered as schemata designating infinite sequences of natural numbers. Given a natural number, we evaluate arithmetic expressions in the following way:

**Definition 36** (Evaluation of arithmetic expressions)**.** Let $a = \alpha \times n + \beta$ be an arithmetic expression, and $N$ a natural number. We define the *evaluation of* $a$ as:

$$a \downarrow N := \alpha \cdot N + \beta \in \mathbb{N}.$$

We define atom schemata, the schematic version of indexed proposition in Definition 2.

**Definition 37** (Atom schema)**.** An atom schema is an expression of the form $P(a)$, where $P$ is a propositional symbol, and $a$ is an arithmetic expression.

Propositional schemata are the schematic version of propositional formulas, which we defined in Definition 3.

**Definition 38** (Propositional schema)**.** The set of propositional schemata is inductively defined as follows:

- each atom schema is a propositional schema;
- the propositional schema variable $X^{\mathrm{prop}}$ is a propositional schema;
- if $F$ is a propositional schema, then $\neg F$ is a propositional schema;
- if $F_1$ and $F_2$ are propositional schemata, then $F_1 \vee F_2$, $F_1 \wedge F_2$, and $F_1 \supset F_2$ are propositional schemata;
- $\mathbf{P}(a)$ is a propositional schema, where $a$ is an arithmetic expression, and $\mathbf{P}$ is a *defined* propositional symbol. $\mathbf{P}$ can be defined by a primitive recursive specification of the form:

  $$\begin{aligned} \mathbf{P}(0) \quad &\equiv \mathrm{F}_{\mathrm{base}} \\ \mathbf{P}(n+1) \quad &\equiv \mathrm{F}_{\mathrm{rec}} \end{aligned}$$

  where $\mathrm{F}_{\mathrm{base}}$ is a propositional formula, and $\mathrm{F}_{\mathrm{rec}}$ is a propositional schema with possible occurrences of $X^{\mathrm{prop}}$. $\mathbf{P}$ cannot occur in $\mathrm{F}_{\mathrm{rec}}$, but can occur other propositional symbols which were previously defined.

  *Note*: we say that $X^{\mathrm{prop}}$ does not occur in the expression $\mathbf{P}(a)$. The intuition is that definitions *bind* the propositional schema variable $X^{\mathrm{prop}}$.

In the following, we will consider only propositional schemata in which the variable $X^{\mathrm{prop}}$ does not occur.

*Remark*: we use bold-case to denote defined propositional symbols, like $\mathbf{P}$ and $\mathbf{Q}$.

**Example 9.** Propositional schemata make it possible to represent generalized conjunctions and disjunctions like the informal expressions $\bigwedge_{i=0}^{n} P(i)$ and $\bigvee_{i=0}^{n} P(i)$.

Let us for example define a propositional schema $\mathbf{Q}(n)$ which represents $\bigwedge_{i=0}^{n} P(i)$. We need the specification:

$$\begin{aligned} \mathbf{Q}(0) \quad &\equiv P(0) \\ \mathbf{Q}(n+1) \quad &\equiv \mathbf{Q}(n) \wedge P(n+1) \end{aligned}$$

which is just syntactic sugar for:

$$\begin{aligned} \mathbf{Q}(0) \quad &\equiv P(0) \\ \mathbf{Q}(n+1) \quad &\equiv X^{\mathrm{prop}} \wedge P(n+1) \end{aligned}$$

As we see, the variable $X^{\mathrm{prop}}$ stands semantically for "$\mathbf{Q}(n)$", which will be made clear in Definition 41.

**Definition 39** (Replacement of propositional schemata)**.** Let $F$ and $G$ be propositional schemata. We denote the replacement in $F$ of $X^{\text{prop}}$ with $G$ by $F[X^{\text{prop}} \mapsto G]$. It is defined by structural induction on $F$:

- $P(b)[X^{\text{prop}} \mapsto G] := P(b)$, where $P$ is a propositional symbol and $a$ an arithmetic expression;
- $\mathbf{P}(b)[X^{\text{prop}} \mapsto G] := \mathbf{P}(b)$, where $P$ is a defined propositional symbol and $a$ an arithmetic expression;
- $X^{\text{prop}}[X^{\text{prop}} \mapsto G] := G$;
- $(\neg F)[X^{\text{prop}} \mapsto G] := \neg(F[X^{\text{prop}} \mapsto G])$;
- $(F_1 \diamond F_2)[X^{\text{prop}} \mapsto G] := (F_1[X^{\text{prop}} \mapsto G]) \diamond (F_2[X^{\text{prop}} \mapsto G])$, where $F_1$ and $F_2$ are propositional schemata, and $\diamond$ is one of the binary connectives $\wedge$, $\vee$ and $\supset$.

**Definition 40** (Replacement in propositional schemata)**.** Let $F$ be a propositional schema and $a$ an arithmetic term. We define $F[n \mapsto a]$ by structural induction on $F$:

- $P(b)[n \mapsto a] := P(b[n \mapsto a])$, where $P$ is a propositional symbol and $a$ an arithmetic expression;
- $\mathbf{P}(b)[n \mapsto a] := \mathbf{P}(b[n \mapsto a])$, where $P$ is a defined propositional symbol and $a$ an arithmetic expression;
- $(\neg F)[n \mapsto a] := \neg(F[n \mapsto a])$, where $F$ is a propositional schema;
- $(F_1 \diamond F_2)[n \mapsto a] := (F_1[n \mapsto a]) \diamond (F_2[n \mapsto a])$, where $F_1$ and $F_2$ are propositional schemata, and $\diamond$ is one of the binary connectives $\wedge$, $\vee$ and $\supset$.

Let us now make precise the idea that propositional schemata encode infinite sequences of propositional formulas.

**Definition 41** (Evaluation of propositional schemata)**.** Let $\alpha$ a natural number. We define the *evaluation of propositional schemata* to propositional formulas by structural induction:

- $P(a) \downarrow \alpha := P(a \downarrow \alpha)$, where $P$ is a propositional symbol and $a$ an arithmetic expression;
- $(\neg F) \downarrow \alpha := \neg\, (F \downarrow \alpha)$, where $F$ is a propositional schema;
- $(F_1 \diamond F_2) \downarrow \alpha := (F_1 \downarrow \alpha) \diamond (F_2 \downarrow \alpha)$, where $F_1$ and $F_2$ are propositional schemata, and $\diamond$ is one of the binary connectives $\wedge$, $\vee$ and $\supset$.

- To define $\mathbf{P}(a) \downarrow \alpha$, where $\mathbf{P}$ has the primitive recursive specification

$$\begin{array}{ll} \mathbf{P}(0) & \equiv \mathrm{F}_{\mathrm{base}} \\ \mathbf{P}(n+1) & \equiv \mathrm{F}_{\mathrm{rec}} \end{array}$$

  we first define the sequence $\{F_\beta\}_{\beta \in \mathbb{N}}$ of propositional formulas:

$$\begin{array}{ll} F_0 & := \mathrm{F}_{\mathrm{base}} \\ F_{\beta+1} & := \mathrm{F}_{\mathrm{rec}}[X^{\mathrm{prop}} \mapsto F_\beta] \downarrow \beta \end{array}$$

  Finally, we set $\mathbf{P}(a) \downarrow \alpha := F_{a\downarrow\alpha}$.

Remark: the evaluation is undefined for $X^{\mathrm{prop}}$.

A propositional schema thus encodes an infinite sequence of propositional formulas in the sense that by evaluating a propositional schema $F$ with natural numbers, we obtain the sequence

$$F \downarrow 0, \;\; F \downarrow 1, \;\; F \downarrow 2, \;\; F \downarrow 3, \;\; \ldots$$

The following definition will be useful when introducing the schematic sequent calculus in Section 2.3. The relation $\leftrightarrow_{def}$ models the unraveling of the definitions.

**Definition 42** ($\leftrightarrow_{def}$)**.** We define the relation $\leftrightarrow_{def}$ on propositional schemata, which we characterize by the reflexive and symmetric closure of the following rule. For every propositional symbol $\mathbf{P}$ defined by a recursive specification like:

$$\begin{array}{ll} \mathbf{P}(0) & \equiv \mathrm{F}_{\mathrm{base}} \\ \mathbf{P}(n+1) & \equiv \mathrm{F}_{\mathrm{rec}} \end{array}$$

we set:

$$\begin{array}{lll} \mathbf{P}(0) & \leftrightarrow_{def} & \mathrm{F}_{\mathrm{base}} \\ \mathbf{P}(\alpha \times n + (\beta+1)) & \leftrightarrow_{def} & \mathrm{F}_{\mathrm{rec}}\,[n \mapsto \alpha \times n + \beta]\,[X^{\mathrm{prop}} \mapsto \mathbf{P}(\alpha \times n + \beta)] \end{array}$$

**Example 10.** Suppose $\mathbf{Q}$ is defined as in Example 9. Then we have:

$$\mathbf{Q}(0) \leftrightarrow_{def} P(0),$$

$$\mathbf{Q}(n+1) \leftrightarrow_{def} \mathbf{Q}(n) \wedge P(n+1).$$

**Proposition 2** (Correctness of $\leftrightarrow_{def}$)**.** If $F_1 \leftrightarrow_{def} F_2$, then for every natural number $\alpha$: $(F_1 \downarrow \alpha) = (F_2 \downarrow \alpha)$.

*Proof.* We need to prove the following: for every natural number $N$

$$\begin{array}{lll} \mathbf{P}(0) \downarrow N & = & \mathrm{F}_{\mathrm{base}} \downarrow N \\ \mathbf{P}(\alpha \times n + (\beta+1)) \downarrow N & = & \mathrm{F}_{\mathrm{rec}}\,[n \mapsto \alpha \times n + \beta]\,[X^{\mathrm{prop}} \mapsto \mathbf{P}(\alpha \times n + \beta)] \downarrow N \end{array}$$

The first equality is easy, since by Definition 41:

$$\mathbf{P}(0) \downarrow N = \mathrm{F}_{\mathrm{base}} = \mathrm{F}_{\mathrm{base}} \downarrow N.$$

The second equality follows similarly from Definition 41 and Definition 40. □

### 2.2.2 Semantics

**Definition 43** (Interpretation)**.** An interpretation of the schematic language is a pair $\langle \mathcal{I}, \alpha \rangle$ of a propositional interpretation $\mathcal{I}$ together with a natural number $\alpha$.

**Definition 44** (Truth value)**.** The truth value $\mathcal{M}(F)$ of a propositional schema $F$ in an interpretation $\mathcal{M} = \langle \mathcal{I}, \alpha \rangle$ is defined as:

$$\mathcal{M}(F) = \mathcal{I}(F \downarrow \alpha)$$

for every propositional schema $F$.

**Definition 45** (Model)**.** Let $F$ be a propositional schema, and $\mathcal{M}$ be an interpretation. We call $\mathcal{M}$ a model of $F$ if and only if $\mathcal{M}(F) = \text{true}$.

We denote that $\mathcal{M}$ is a model of $F$ by $\mathcal{M} \models F$.

**Definition 46** (Satisfiability)**.** Let $F$ be a propositional schema.

- $F$ is called *satisfiable* if $F$ has a model.
- $F$ is called *unsatisfiable* if it is not satisfiable.
- $F$ is called *valid* if every interpretation is a model of $F$.

We conclude this section with a remark on the satisfiability of propositional schemata. The satisfiability problem for propositional logic (the *SAT problem*) is well-known, and the most publicized problem in the *NP class*, defined in complexity theory as the set of decision problems solvable in polynomial time by a theoretical non-deterministic Turing machine. Unfortunately, the corresponding satisfiability problem for propositional schemata is beyond computability:

**Theorem 2.** The set of unsatisfiable propositional schemata is not recursively enumerable.

*Proof.* This is a consequence of Theorem 6.2 in [ACP11], since our notion of schemata is a superset of their notion of “homothetic schemata”, whose satisfiability is proven undecidable. □

## 2.3 Schematic Sequent Calculus

We now introduce a sequent calculus for propositional schemata. Let us generalize the definitions in Section 1.2 to the schematic case.

**Definition 47** (Sequent schema)**.** Let $\Gamma$ and $\Delta$ be lists of propositional schemata. The expression $\Gamma \vdash \Delta$ is called a *sequent schema.* “$\vdash$” is called the *empty sequent schema.*

**Definition 48** (Substitution in sequent schemata)**.** Let $S$ be a sequent schema and $a$ an arithmetic term. We denote by $S[n \mapsto a]$ the sequent schema obtained intuitively by replacing each propositional schema $F$ in $S$ with $F[n \mapsto a]$.

As before, sequent schemata represent infinite sequences of sequents, which can be made explicit by evaluating $n$ to a natural number:

**Definition 49** (Evaluation of sequent schemata)**.** Let $N$ be a natural number; the *evaluation of a sequent schemata* is defined by:

$$(F_1, \ldots, F_\alpha \vdash G_1, \ldots, G_\beta) \downarrow N := (F_1 \downarrow N), \ldots, (F_\alpha \downarrow N) \vdash (G_1 \downarrow N), \ldots, (G_\beta \downarrow N).$$

Similarly as in Definition 14, we define:

**Definition 50** (Initial sequent schema)**.** An *initial sequent schema* is a sequent schema of the form $P(a) \vdash P(a)$, where $P(a)$ is an atom schema.

Before defining *proof schemata*, let us define the unraveling of definitions in sequents:

**Definition 51** (Definitions)**.** We extend $\leftrightarrow_{def}$ from Definition 42 to sequent schemata in the following way:

$$(F_1, \ldots, F_\alpha \vdash G_1, \ldots, G_\beta) \leftrightarrow_{def} (F'_1, \ldots, F'_\alpha \vdash G'_1, \ldots, G'_\beta)$$

if and only if $F_1 \leftrightarrow_{def} F'_1$, ..., $F_\alpha \leftrightarrow_{def} F'_\alpha$ and $G_1 \leftrightarrow_{def} G'_1$, ..., $G_\beta \leftrightarrow_{def} G'_\beta$.

We are now ready to define the schematic version of propositional proofs.

**Definition 52** (Proof symbol)**.** We assume a countably infinite set of proof symbols, which we denote by abuse of notation with $\varphi$, $\psi$, etc.

Proof links act like placeholders for proof schemata. Similar approaches may be found on the literature on *cyclic proofs* (see [Bro05] and [SD03]).

**Definition 53** (Proof Link)**.** A *proof link* is an expression of the form:

$$\frac{\boldsymbol{\varphi}(a)}{S}$$

where $\varphi$ is a proof symbol, $a$ an arithmetic term, and $S$ a sequent schema.

We define inferences for propositional schemata, similarly as in Definition 15.

**Definition 54** (Inference)**.** Inferences for the schematic sequent calculus are defined as in Definition 15: there are rules equivalent to $w$, $c$, $x$, $\wedge$, $\vee$, $\supset$, $\neg$ (in the left and right variants), and *cut*. In addition, the schematic sequent calculus features the following rules of inference:

- Definition rule:

$$\frac{S}{S'}\,def$$

where $S \leftrightarrow_{def} S'$.

- Proof link rule: every proof link is an inference

$$\frac{\varphi(a)}{S}.$$

We define proof schemata in a similar way as propositional proofs in Definition 16:

**Definition 55** (Proof schemata)**.** We denote a proof schema $\varphi$ with end-sequent $S$ by:

$$\begin{array}{c}\varphi\\ \vdots\\ S\end{array} \quad \text{or} \quad \begin{array}{c}(\varphi)\\ S\end{array} \quad \text{or simply} \quad \varphi\ .$$

The set of proof schemata is inductively defined as follows:

- an initial sequent schema $S$ is a proof schema with end-sequent $S$;
- if $\varphi'$ is a proof schema with end-sequent $S'$, and $\frac{S'}{S}$ is a unary inference or a definition inference, then

$$\frac{\begin{array}{c}\varphi'\\ \vdots\\ S'\end{array}}{S}$$

is a proof schema with end-sequent $S$;
- if $\varphi_1$ is a proof schema with end-sequent $S_1$, $\varphi_2$ is a proof schema with end-sequent $S_2$, and $\frac{S_1 \quad S_2}{S}$ is a binary inference, then

$$\frac{\begin{array}{c}\varphi_1\\ \vdots\\ S_1\end{array} \quad \begin{array}{c}\varphi_2\\ \vdots\\ S_2\end{array}}{S}$$

is a proof schema with end-sequent $S$.

- $\boldsymbol{\varphi}(a)$ is a proof schema, where $a$ is an arithmetical term, and $\boldsymbol{\varphi}$ is a proof symbol defined by primitive recursion, by providing the rules:

$$\boldsymbol{\varphi}(0) \equiv \begin{array}{c}\varphi_{\text{base}} \\ \vdots \\ S \downarrow 0\end{array}$$

$$\boldsymbol{\varphi}(n+1) \equiv \begin{array}{c}\varphi_{\text{rec}} \\ \vdots \\ S[n \mapsto n+1]\end{array}$$

  such that:

  - $\varphi_{\text{base}}$ is a propositional proof with end-sequent $S \downarrow 0$, and
  - $\varphi_{\text{rec}}$ is a proof schema with end-sequent $S[n \mapsto n+1]$ which may contain only proof links of the form $\dfrac{\boldsymbol{\varphi}(n)}{S}$.

  Then $\boldsymbol{\varphi}(a)$ is a proof schema with end-sequent $S[n \mapsto a]$.

  *Note*: we say that no proof link occurs in $\boldsymbol{\varphi}(a)$. The intuition is that the primitive recursive specification *binds* proof links.

In the following, we will consider only proof schemata in which there are no occurrences of proof links.

*Remark*: we use bold-case to denote defined proof symbols, like $\boldsymbol{\varphi}$ and $\boldsymbol{\psi}$.

**Example 11.** For examples of proof schemata, see Chapter 6.

**Definition 56** (Replacement of proofs)**.** Let $\varphi_1$ and $\varphi_2$ be proof schemata, where in $\varphi_1$ there are proof links. The replacement of the proof links in $\varphi_1$ with $\varphi_2$ can be defined as one would imagine, by structural induction on $\varphi_1$.

**Definition 57** (Evaluation of proof schemata)**.** Let $\alpha$ be a natural number. We define the *evaluation of proof schemata* to propositional proofs by structural induction:

- If $\varphi$ is the initial sequent schema $S$, then $\varphi \downarrow \alpha := S \downarrow \alpha$;
- If $\varphi$ ends with a unary or binary propositional inference, then the evaluation is defined recursively on the corresponding sub-proofs as one would expect;
- If $\varphi$ ends with a definition inference:

$$\dfrac{\begin{array}{c}(\varphi') \\ S\end{array}}{S'}\, def$$

then $\varphi \downarrow \alpha := \varphi' \downarrow \alpha$;

- To define $\boldsymbol{\varphi}(a) \downarrow \alpha$, where $\boldsymbol{\varphi}$ has the primitive recursive specification

$$\begin{aligned} \boldsymbol{\varphi}(0) &\equiv \varphi_{\text{base}} \\ \boldsymbol{\varphi}(n+1) &\equiv \varphi_{\text{rec}} \end{aligned}$$

we first define the sequence $\{\varphi_\beta\}_{\beta\in\mathbb{N}}$ of propositional proofs:

$$\begin{aligned} \varphi_0 &:= \varphi_{\text{base}} \\ \varphi_{\beta+1} &:= (\text{replace proof links in } \varphi_{\text{rec}} \text{ with } \varphi_\beta) \downarrow \beta \end{aligned}$$

Finally, we set $\boldsymbol{\varphi}(a) \downarrow \alpha := \varphi_{a\downarrow\alpha}$.

Remark: the evaluation is undefined for proof links.

The notions of *configuration* and *induced configuration* (Definitions 18 and 19) are easily adapted to the schematic case.

## 2.4 Schematic Resolution Calculus

We define the schematic version of the resolution calculus in Section 1.3. The central notion here is that of *resolution proof schemata* (Definition 59), which represent infinite sequences of resolution derivations.

We define *clause schemata* following Definition 20 for clauses.

**Definition 58** (Clause schema)**.** A *clause schema* is a sequent schema which contains only atom schemata.

*Remark*: As we will see in Chapter 4, the above definition of clause schemata is too weak, and it will we reformulated in Definition 73.

Similarly as in Definition 21, a clause schema which contains an atom schema both in its antecedent and succedent is called a *tautology.*

We assume a countably infinite set of resolution proof schema symbols, which we denote by abuse of notation with $\rho$, possibly modified with sub- or super-scripts.

We follow Definition 24 and generalize the concept of resolution deductions to *resolution proof schemata.*

**Definition 59** (Resolution proof schema)**.** We define inductively the set of *resolution proof schemata* by:

- a clause schema is a resolution proof schema;

- the variable $X^{\text{res}}$ is a resolution proof schema;
- if $\rho_1$ and $\rho_2$ are resolution proof schemata, and $A$ is an atom schema, then $\mathrm{r}(\rho_1, \rho_2; A)$ is a resolution proof schema;
- $\boldsymbol{\rho}(a)$ is a resolution proof schema, where $a$ is an arithmetic expression, and $\boldsymbol{\rho}$ is a resolution proof schema defined by a primitive recursive specification of the form:

$$\begin{array}{lcl} \boldsymbol{\rho}(0) & \equiv & \rho_{\text{base}} \\ \boldsymbol{\rho}(n+1) & \equiv & \rho_{\text{rec}}[X^{\text{res}}] \end{array}$$

  where $\rho_{\text{base}}$ is a resolution derivation, and $\rho_{\text{rec}}$ is a resolution proof schema in which can occur the resolution proof schema variable $X^{\text{res}}$.

  *Note*: we say that $X^{\text{res}}$ does not occur in $\boldsymbol{\rho}(a)$. The intuition is that the primitive recursive specification *binds* the variable.

In the following, we will consider only resolution proof schemata in which there are no occurrences of $X^{\text{res}}$.

*Remark*: we use bold-case to denote defined resolution proof schema symbols, like $\boldsymbol{\rho}$.

**Definition 60** (Replacement of resolution proof schemata)**.** Let $\rho$ and $\rho'$ resolution proof schemata. We denote with $\rho[X^{\text{res}} \mapsto \rho']$ the schema obtained by replacing the variable $X^{\text{res}}$ in $\rho$ with $\rho'$, and it is defined (as usual) by structural induction on $\rho$:

- $C[X^{\text{res}} \mapsto \rho'] := C$;
- $\mathrm{r}(\rho_1, \rho_2; A)[X^{\text{res}} \mapsto \rho'] := \mathrm{r}(\rho_1[X^{\text{res}} \mapsto \rho'], \rho_2[X^{\text{res}} \mapsto \rho']; A)$
- $X^{\text{res}}[X^{\text{res}} \mapsto \rho'] := \rho'$;
- $\boldsymbol{\rho}(a)[X^{\text{res}} \mapsto \rho'] := \boldsymbol{\rho}(a)$, where $\boldsymbol{\rho}$ is a defined resolution proof schema symbol.

**Definition 61** (Evaluation of resolution proof schemata)**.** Let $\alpha$ be a natural number. We define the *evaluation of* resolution proof schemata to resolution derivations, by structural recursion:

- if $\rho$ is the clause schema $C$, then $\rho \downarrow \alpha := C \downarrow \alpha$;
- $\mathrm{r}(\rho_1, \rho_2; A) \downarrow \alpha := \mathrm{r}(\rho_1 \downarrow \alpha, \rho_2 \downarrow \alpha; A \downarrow \alpha)$;
- To define $\boldsymbol{\rho}(a) \downarrow \alpha$, where $\boldsymbol{\rho}$ has the primitive recursive specification

$$\begin{array}{ll} \boldsymbol{\rho}(0) & \equiv \rho_{\text{base}} \\ \boldsymbol{\rho}(n+1) & \equiv \rho_{\text{rec}} \end{array}$$

we first define the sequence $\{\rho_\beta\}_{\beta\in\mathbb{N}}$ of resolution derivations:

$$\begin{aligned}\rho_0 &:= \rho_{\text{base}}\\ \rho_{\beta+1} &:= (\rho_{\text{rec}}[X^{\text{res}} \mapsto \rho_\beta]) \downarrow \beta\end{aligned}$$

Finally, we set $\boldsymbol{\rho}(a) \downarrow \alpha := \rho_{a\downarrow\alpha}$.

Remark: the evaluation is undefined for the variable $X^{\text{res}}$.

**Definition 62** (Resolution refutation schema)**.** A *resolution refutation schema* $\rho$ is a resolution proof schema such that for every natural number $\alpha$, $\rho \downarrow \alpha$ is a resolution refutation.

**Example 12.** Let us define the following resolution derivation schema:

$$\rho := \text{r}((P(n+1) \vdash P(n)), (P(n) \vdash); P(n))$$

$\rho$ is a resolution derivation schema for the clause schema $P(n+1) \vdash$: for every $\alpha$, $\rho \downarrow \alpha$ is a resolution deduction with end-clause $P(\alpha+1) \vdash$.

A schematic version of the resolution calculus with weakenings in Section 1.3.1 is not necessary for now, but it could be investigated in the future.

## 2.5 Ceres for Propositional Schemata

In this section, we are going to adapt Ceres to the case of propositional schemata. This requires adapting the concepts of *clause set term*, *characteristic term*, and *proof projection* to proof schemata. The result will be the concepts of *clause set term schema*, *characteristic term schema*, and *proof projection schema.*

Let us start by porting Definition 28 to the schematic case. The following definition is slightly different than the previous ones in this chapter: we now allow a more expressive way of defining symbols for clause set term schemata in a recursive way. The previous schemata allowed only primitive recursive definitions, while for clause set term schemata we must allow *mutual recursive definitions.* This means that we allow to define multiple clause set term schemata at once, where each one of them can depend recursively on itself and on the others.

We assume a countably infinite set of clause set term schema symbols, which we denote by abuse of notation with $T$ with possible sub- and super-scripts.

**Definition 63** (Clause set term schema)**.** *Clause set term schemata* are defined inductively by:

- A clause set term schema variable is a clause term schema. We assume a countably infinite set of clause set term schema variables which we denote by $X_1^{\text{term}}, X_2^{\text{term}}, \ldots$;

- $[C]$ is a clause set term schema, where $C$ is a clause schema;
- $T_1 \oplus T_2$ and $T_1 \otimes T_2$ are clause set term schemata, where $T_1$ and $T_2$ are clause set term schemata;
- $\mathbf{T}(a)$ is a clause set term schema, where $a$ is an arithmetic expression and $\mathbf{T}$ is defined by mutual recursion in the following way. Let $\mathbf{T}^1, \ldots, \mathbf{T}^N$ be $N$ clause set term schema symbols: we provide a mutual recursive definition by means of the following equations:
$$\begin{array}{ll} \mathbf{T}^1(0) & \equiv \mathrm{T}^1_{\text{base}} \\ \mathbf{T}^1(n+1) & \equiv \mathrm{T}^1_{\text{rec}} \\ \vdots & \vdots \\ \mathbf{T}^N(0) & \equiv \mathrm{T}^N_{\text{base}} \\ \mathbf{T}^N(n+1) & \equiv \mathrm{T}^N_{\text{rec}} \end{array}$$
where $\mathrm{T}^1_{\text{base}}, \ldots, \mathrm{T}^N_{\text{base}}$ are clause set terms, and $\mathrm{T}^1_{\text{rec}}, \ldots, \mathrm{T}^N_{\text{rec}}$ are clause set term schemata which can have occurrences of the variables $X^{\text{term}}_1, \ldots, X^{\text{term}}_N$.

  *Note*: we say that the variables like $X^{\text{term}}$ do not occur in $\boldsymbol{T}(a)$. The intuition is that the recursive specification *binds* variables.

In the following, we will consider only clause set term schemata in which there are no occurrences of clause set term schema variables.

*Note*: we will denote defined clause set term schema symbols in bold-case, like $\mathbf{T}$.

**Definition 64** (Replacement of clause set term schema variables)**.** Let $T$ and $T'$ clause set term schemata. We denote with $T[X^{\text{term}}_H \mapsto T']$ the schema obtained by replacing the variable $X^{\text{term}}_H$ in $T$ with $T'$, and it is defined (as usual) by structural induction on $T$:

- $[C][X^{\text{term}}_H \mapsto T'] := [C]$;
- $(T_1 \otimes T_2)[X^{\text{term}}_H \mapsto T'] := (T_1[X^{\text{term}}_H \mapsto T']) \otimes (T_2[X^{\text{term}}_H \mapsto T'])$, and simiarly for $\oplus$;
- $X^{\text{term}}_H[X^{\text{term}}_H \mapsto T'] := T'$;
- $X^{\text{term}}_K[X^{\text{term}}_H \mapsto T'] := X^{\text{term}}_K$ for $K \neq H$;
- $\boldsymbol{T}(a)[X^{\text{term}}_H \mapsto T'] := \boldsymbol{T}(a)$, where $\boldsymbol{T}$ is a defined clause set term schema symbol.

**Definition 65** (Evaluation of clause set term schemata)**.** Let $T$ be a clause set term schema, and $\alpha$ a natural number. We define the *evaluation of* $T$ to a clause set term, by structural induction on $T$:

- if $T$ is $[C]$ for a clause schema $C$, then $T \downarrow \alpha := [C \downarrow \alpha]$;
- $(T_1 \oplus T_2) \downarrow \alpha := (T_1 \downarrow \alpha) \oplus (T_2 \downarrow \alpha)$, and similarly for the case of $\otimes$;

- To define $\boldsymbol{T}(a) \downarrow \alpha$, where $\boldsymbol{T} = \mathbf{T}^H$ was defined by the following mutual recursive specification:

$$\begin{array}{ll} \mathbf{T}^1(0) & \equiv \mathrm{T}^1_{\text{base}} \\ \mathbf{T}^1(n+1) & \equiv \mathrm{T}^1_{\text{rec}} \\ \vdots & \vdots \\ \mathbf{T}^N(0) & \equiv \mathrm{T}^N_{\text{base}} \\ \mathbf{T}^N(n+1) & \equiv \mathrm{T}^N_{\text{rec}} \end{array}$$

  we first define the sequences $\{T_{\beta,K}\}_{\beta\in\mathbb{N}}$ of clause set terms, for $K = 1 \dots N$:

$$\begin{array}{ll} T_{0,K} & := \mathrm{T}^K_{\text{base}} \\ T_{\beta+1,K} & := (\mathrm{T}^K_{\text{rec}}[X_1^{\text{term}} \mapsto T_{\beta,1}] \cdots [X_N^{\text{term}} \mapsto T_{\beta,N}]) \downarrow \beta \end{array}$$

  Finally, we set $\boldsymbol{T}(a) \downarrow \alpha := T_{a\downarrow\alpha,H}$.

Remark: the evaluation is undefined for the variables $X^{\text{term}}$.

**Example 13.** Let us give the following mutual recursive definition:

$$\begin{array}{ll} \mathbf{T}_1(0) & \equiv [\vdash] \\ \mathbf{T}_1(n+1) & \equiv [P(n) \vdash] \otimes \mathbf{T}_2(n) \\ \mathbf{T}_2(0) & \equiv [\vdash] \\ \mathbf{T}_2(n+1) & \equiv [\vdash Q(n)] \oplus \mathbf{T}_1(n) \end{array}$$

which is just syntactic sugar for:

$$\begin{array}{ll} \mathbf{T}_1(0) & \equiv [\vdash] \\ \mathbf{T}_1(n+1) & \equiv [P(n) \vdash] \otimes X_2^{\text{term}} \\ \mathbf{T}_2(0) & \equiv [\vdash] \\ \mathbf{T}_2(n+1) & \equiv [\vdash Q(n)] \oplus X_1^{\text{term}} \end{array}$$

We have:

$$\begin{array}{ll} \mathbf{T}_1 \downarrow 1 & = [P(0) \vdash] \otimes [\vdash], \\ \mathbf{T}_2 \downarrow 1 & = [\vdash Q(0)] \oplus [\vdash], \\ \mathbf{T}_1 \downarrow 2 & = [P(1) \vdash] \otimes ([\vdash Q(0)] \oplus [\vdash]). \\ \vdots & \vdots \end{array}$$

We give Definition 66 by adapting Definition 30 to proof schemata:

**Definition 66** (Characteristic term schema)**.** Let $\varphi$ be a proof schema with end-sequent $S$, and $\Omega$ a configuration for $S$. We define by structural recursion on $\varphi$:

- If $\varphi$ consists of the initial sequent schema $S$, then $\Theta^{\Omega}(\varphi) := [S \cdot \Omega]$;
- If $\varphi$ ends with a unary inference:

$$\dfrac{\begin{matrix}(\varphi')\\ S'\end{matrix}}{S}$$

then $\Theta^{\Omega}(\varphi) := \Theta^{\Omega'}(\varphi')$ where $\Omega'$ is the configuration induced by $\Omega$ on $S'$.

- If $\varphi$ ends with a binary inference:

$$\dfrac{\begin{matrix}(\varphi_1)\\ S_1\end{matrix} \qquad \begin{matrix}(\varphi_2)\\ S_2\end{matrix}}{S}$$

let $\Omega_1$ and $\Omega_2$ be the configurations induced by $\Omega$ respectively on $S_1$ and $S_2$:

  - if the principal formula of the inference is in $\Omega$, or if the inference is a cut, then $\Theta^{\Omega}(\varphi) := \Theta^{\Omega_1}(\varphi_1) \oplus \Theta^{\Omega_2}(\varphi_2)$;
  - otherwise then $\Theta^{\Omega}(\varphi) := \Theta^{\Omega_1}(\varphi_1) \otimes \Theta^{\Omega_2}(\varphi_2)$.

- If $\varphi$ is a proof link:

$$\begin{matrix}\boldsymbol{\psi}(n)\\ \cdots\cdots\\ S\end{matrix}$$

then $\Theta^{\Omega}(\varphi) := X_{\Omega}^{\text{term}}$. The intended semantics of $X_{\Omega}^{\text{term}}$ is "$\Theta^{\Omega}(\boldsymbol{\psi})(n)$";

- If $\varphi$ is $\boldsymbol{\psi}(a)$ where $\boldsymbol{\psi}$ has the recursive specification:

$$\begin{array}{lcl}\boldsymbol{\psi}(0) & \equiv & \psi_{\text{base}}\\ \boldsymbol{\psi}(n+1) & \equiv & \psi_{\text{rec}}\end{array}$$

Then we first define $\Theta^{\Omega_1}(\boldsymbol{\psi}), \ldots, \Theta^{\Omega_\alpha}(\boldsymbol{\psi})$ for all the configurations $\Omega_1, \ldots, \Omega_\alpha$ for the end-sequent schema $S$ at the same time, by mutual recursion:

$$\begin{array}{ll}\Theta^{\Omega_1}(\boldsymbol{\psi})(0) & \equiv \Theta^{\Omega_1}(\psi_{\text{base}})\\ \Theta^{\Omega_1}(\boldsymbol{\psi})(n+1) & \equiv \Theta^{\Omega_1}(\psi_{\text{rec}})\\ \vdots & \quad\vdots\\ \Theta^{\Omega_\alpha}(\boldsymbol{\psi})(0) & \equiv \Theta^{\Omega_\alpha}(\psi_{\text{base}})\\ \Theta^{\Omega_\alpha}(\boldsymbol{\psi})(n+1) & \equiv \Theta^{\Omega_\alpha}(\psi_{\text{rec}})\end{array}$$

Then we set $\Theta^{\Omega}(\boldsymbol{\psi}(a)) := \Theta^{\Omega}(\boldsymbol{\psi})(a)$.

The other important steps of Ceres are:

- computing the *proof projections* (which in the schematic case become *proof projection schemata*), and
- finally merging the resolution refutation of the characteristic term with the proof projections, thus obtaining a proof with only atomic cuts.

As we already noted in Section 1.4.2, the main topic of this thesis is exploring different formalisms for resolution proof schemata, which should allow to refute every characteristic term schema. Therefore, we focused on the study of characteristic term schemata only, leaving proof projections and the following steps of Ceres for the future.

However, projection schemata can be easily defined, in a similar way as we did in Definition 66, following [DLRW13]. Concerning the final merging of resolution refutation schema and projection schemata, the problem of finding a purely primitive recursive schematic formalism is still open.

CHAPTER 3

# Generic Clause Sets

The first step towards a complete Ceres for propositional schemata is finding uniformity in the clause sets extracted from proofs. In fact, the structure of characteristic clause sets can be quite nested and complicated. Since our goal is to refute characteristic term schemata of proof schemata *in a uniform way*, it would be very helpful if every characteristic clause set could be standardized to a generic clause set, which does not depend on the particular proof schema, but only on the cut configuration, cut formulas, or atoms occurring in it.

In this chapter, we are going to develop two forms of *generic* clause sets; generic, in the sense that they have a simple specification, and also because other clause sets – no matter how complicated – can be reduced to these regular structures:

1. *top clause sets*, which are used as archetypal unsatisfiable sets of clauses;
2. *canonic characteristic clause sets*, which are used as archetypal characteristic clause sets, depending on the configuration $\Omega$.

## 3.1 Top Clause Sets

**Definition 67** (Top Clause Set)**.** Let $A$ be a propositional atom, and $\mathcal{A}$ and $\mathcal{A}'$ be multisets (Definition 8) of propositional atoms. We define the operator $\mathrm{CL}^t(\cdot)$ which maps a multiset of propositional atoms to a clause set:

$$\begin{aligned} \mathrm{CL}^t(\wr A \wr) &:= \{A \vdash;\ \vdash A\} \\ \mathrm{CL}^t(\mathcal{A} \,\dot{\cup}\, \mathcal{A}') &= \mathrm{CL}^t(\mathcal{A}) \times \mathrm{CL}^t(\mathcal{A}') \end{aligned}$$

where the product $\times$ of clause sets was introduced in Definition 29.

We call $\mathrm{CL}^t(\mathcal{A})$ the *top clause set* with respect to the atoms in $\mathcal{A}$.

Note: it follows from the definition above that $\mathrm{CL}^t\left(\lbag\rbag\right) = \{\vdash\}$.

Let us first prove a semantic lemma which will help to prove Proposition 3 below.

**Lemma 1.** Let $S_1$ and $S_2$ be clause sets. If $S_1$ and $S_2$ are unsatisfiable, then $S_1 \times S_2$ is unsatisfiable too.

*Proof.* Note that $S_1 \times S_2$ is just the conjunctive normal form of the disjunction of $S_1$ and $S_2$. Both $S_1$ and $S_2$ are unsatisfiable, therefore their disjunction is unsatisfiable too. □

**Proposition 3.** Every top clause set is unsatisfiable.

*Proof.* Let $\mathcal{A}$ be a multiset of propositional atoms. Let us proceed by induction on the number of atoms in that set of atoms:

- If $\mathcal{A}$ is empty, then $\mathrm{CL}^t\left(\lbag\rbag\right) = \{\vdash\}$, which is clearly unsatisfiable.
- If $\mathcal{A}$ has just one element, then we conclude since again $\mathrm{CL}^t\left(\lbag A\rbag\right) = \{A \vdash;\ \vdash A\}$ is clearly unsatisfiable.
- Suppose now $\mathcal{A}$ has more than one element. Then we can partition the multiset in two non-empty sub-multisets $\mathcal{A} = \mathcal{A}' \mathbin{\dot\cup} \mathcal{A}''$. By inductive hypothesis, both $\mathrm{CL}^t\left(\mathcal{A}'\right)$ and $\mathrm{CL}^t\left(\mathcal{A}''\right)$ are unsatisfiable. By Lemma 1, we conclude that $\mathrm{CL}^t\left(\mathcal{A}'\right) \times \mathrm{CL}^t\left(\mathcal{A}''\right)$ is unsatisfiable too, which is exactly $\mathrm{CL}^t\left(\mathcal{A}\right)$.

□

Let us now study the shape of top clause sets: Lemma 2 shows that top clause sets are actually characteristic clause sets of particular proofs.

**Lemma 2.** For every multiset $\mathcal{A}$ of propositional atoms, there is a proof $\varphi$ containing only atomic cuts at the top of the proof, whose cut-formulas are exactly the ones in $\mathcal{A}$ (with the right multiplicities), such that $\mathrm{CL}(\varphi) = \mathrm{CL}^t\left(\mathcal{A}\right)$.

*Proof.* It is easy to construct such a proof $\varphi$, given a multiset of atoms $\mathcal{A}$. For example, let us consider $\mathcal{A} = \lbag A_1, A_2, A_3\rbag$, and the following proof $\varphi$:

$$\dfrac{\dfrac{\dfrac{\dfrac{A_1 \vdash A_1 \quad A_1 \vdash A_1}{A_1 \vdash A_1}\,\text{cut}}{A_1 \wedge A_2 \wedge A_3 \vdash A_1}\,\wedge\colon l \quad \dfrac{\dfrac{A_2 \vdash A_2 \quad A_2 \vdash A_2}{A_2 \vdash A_2}\,\text{cut}}{A_1 \wedge A_2 \wedge A_3 \vdash A_2}\,\wedge\colon l}{A_1 \wedge A_2 \wedge A_3 \vdash A_1 \wedge A_2} \quad \dfrac{\dfrac{A_3 \vdash A_3 \quad A_3 \vdash A_3}{A_3 \vdash A_3}\,\text{cut}}{A_1 \wedge A_2 \wedge A_3 \vdash A_3}\,\wedge\colon l}{A_1 \wedge A_2 \wedge A_3 \vdash A_1 \wedge A_2 \wedge A_3}$$

It is easy to see that the characteristic clause set of $\varphi$ is exactly the top clause set of $\wr A_1, A_2, A_3 \wr$. $\square$

Until this point, we only considered top clause sets in the context of propositional logic. In Chapter 5, we are going to generalize these results to the case of propositional schemata. Multisets of atoms will not suffice anymore, and it will be necessary to introduce the notion of *atom set schemata.*

For now, let us consider the following example, which shows how to generalize to schemata the proof of Lemma 2:

**Example 14.** In the proof for Lemma 2, we saw how to construct propositional proofs which have top clause sets as characteristic clause sets.

In this example, we want to generalize the result to an arbitrary and increasing number of propositional atoms. In order to do so, we define a proof schema with an increasing number of atomic cuts. This will suggest the way to extend top clause sets to the case of propositional schemata.

The key in Lemma 2 was to construct conjunctions of atoms; the generalization to schemata yields the following generalized conjunctions:

$$\begin{aligned} \mathbf{Q}(0) &\equiv P(0) \\ \mathbf{Q}(n+1) &\equiv \mathbf{Q}(n) \wedge P(n+1) \end{aligned}$$

Remark: $\mathbf{Q}(n)$ defines $\bigwedge_{i=0}^{n} P(i)$.

Let $\boldsymbol{\varphi}$ the following proof schema:

**Proof schema:** $\boldsymbol{\varphi}$

$$\boldsymbol{\varphi}(0) \quad \equiv \quad \dfrac{\dfrac{P(0) \vdash P(0) \qquad P(0) \vdash P(0)}{P(0) \vdash P(0)}\,\text{cut}}{\mathbf{Q}(0) \vdash \mathbf{Q}(0)}\,\text{def} \tag{3.1}$$

$$\boldsymbol{\varphi}(n+1) \quad \equiv \quad \text{see below}$$

$$\dfrac{\dfrac{\dfrac{\dfrac{[\boldsymbol{\varphi}(n)]}{\mathbf{Q}(n) \vdash \mathbf{Q}(n)}}{\mathbf{Q}(n) \wedge P(n+1) \vdash \mathbf{Q}(n)}\,\wedge\colon l \qquad \dfrac{\dfrac{P(n+1) \vdash P(n+1) \qquad P(n+1) \vdash P(n+1)}{P(n+1) \vdash P(n+1)}\,\text{cut}}{\mathbf{Q}(n) \wedge P(n+1) \vdash P(n+1)}\,\wedge\colon l}{\mathbf{Q}(n) \wedge P(n+1) \vdash \mathbf{Q}(n) \wedge P(n+1)}\,\wedge\colon r}{\mathbf{Q}(n+1) \vdash \mathbf{Q}(n+1)}\,\text{def}$$

For every natural number $\alpha$, we have:

$$|\Theta\left(\boldsymbol{\varphi}(n)\right) \downarrow \alpha| = \mathrm{CL}^t\left(\{P(0), \ldots, P(\alpha)\}\right).$$

For example:

$$\begin{array}{ll}\mathrm{CL}^t(\{P(0)\}) & = \{P(0) \vdash;\ \vdash P(0)\} \\ \mathrm{CL}^t(\{P(0), P(1)\}) & = \{P(0), P(1) \vdash;\ P(0) \vdash P(1); \\ & \quad \vdash P(0), P(1);\ P(1) \vdash P(0)\} \\ & \vdots \end{array}$$

## 3.2 Canonic Characteristic Clause Sets

In this section, we are going to define what we call *canonic characteristic clause sets*. These clause sets are interesting because they are archetypal characteristic clause sets, in the sense that they are the *least clause set logically entailed by all characteristic clause sets* with the same end-sequent and on the same cut configuration.

**Definition 68** (Canonic characteristic clause set)**.** We map every formula to its associated left (resp. right) clause set:

$$\begin{array}{llll}\mathcal{L}(A) & := \{A \vdash\} & \mathcal{R}(A) & := \{\vdash A\} \\ \mathcal{L}(\neg F) & := \mathcal{R}(F) & \mathcal{R}(\neg F) & := \mathcal{L}(F) \\ \mathcal{L}(F \wedge G) & := \mathcal{L}(F) \times \mathcal{L}(G) & \mathcal{R}(F \wedge G) & := \mathcal{R}(F) \cup \mathcal{R}(G) \\ \mathcal{L}(F \vee G) & := \mathcal{L}(F) \cup \mathcal{L}(G) & \mathcal{R}(F \vee G) & := \mathcal{R}(F) \times \mathcal{R}(G) \\ \mathcal{L}(F \to G) & := \mathcal{R}(F) \cup \mathcal{L}(G) & \mathcal{R}(F \to G) & := \mathcal{L}(F) \times \mathcal{R}(G)\end{array}$$

Let now $F_1, \ldots, F_\alpha, G_1, \ldots, G_\beta$ be formulas. We define the canonic clause set of a sequent:

$$\begin{array}{l}\mathcal{C}(\quad \vdash \quad) := \{\vdash\} \\ \mathcal{C}(F_1, \ldots, F_\alpha \vdash G_1, \ldots, G_\beta) := \mathcal{L}(F_1) \times \ldots \times \mathcal{L}(F_\alpha) \times \mathcal{R}(G_1) \times \ldots \times \mathcal{R}(G_\beta).\end{array}$$

The definition above is non-schematic, i.e. it does not take into account formula schemata; but it is easy to extend the definition to the case of propositional schemata.

In Proposition 4, we extend the concept of canonic clause set to formulas, which will be useful in this section and in the following chapters.

**Proposition 4.** For every formula $F$, there exists a proof $\pi$ such that $\mathrm{CL}(\pi) = \mathcal{L}(F) \cup \mathcal{R}(F)$. We denote this clause set by $\mathcal{C}(F) := \mathcal{L}(F) \cup \mathcal{R}(F)$, which we call the *canonic clause set* of the formula $F$.

*Proof.* Given $F$, the proof $\pi$ is:

$$\dfrac{\begin{array}{c}(\pi_F) \\ F \vdash F\end{array} \qquad \begin{array}{c}(\pi_F) \\ F \vdash F\end{array}}{F \vdash F}\ cut$$

where the proof $\pi_F$ can be defined by structural recusion on $F$:

- if $F$ is an atom, then $\pi_F$ is just the initial sequent $F \vdash F$;
- if $F = \neg G$, then $\pi_{\neg G}$ is:

$$\dfrac{\dfrac{\begin{matrix}(\pi_G)\\ G \vdash G\end{matrix}}{\vdash G, \neg G}\ \neg\colon l}{\neg G \vdash \neg G}\ \neg\colon r$$

- if $F = F_1 \wedge F_2$, then $\pi_{F_1 \wedge F_2}$ is:

$$\dfrac{\dfrac{\begin{matrix}(\pi_{F_1})\\ F_1 \vdash F_1\end{matrix}}{F_1 \wedge F_2 \vdash F_1}\ \wedge\colon l_2 \qquad \dfrac{\begin{matrix}(\pi_{F_2})\\ F_2 \vdash F_2\end{matrix}}{F_1 \wedge F_2 \vdash F_1}\ \wedge\colon l_1}{F_1 \wedge F_2 \vdash F_1 \wedge F_2}\ \wedge\colon r$$

- similarly for $F = F_1 \vee F_2$ and $F = F_1 \supset F_2$.

It is easy to see that $\mathrm{CL}(\pi_F) = \mathcal{L}(F) \cup \mathcal{R}(F)$.

In particular, the following equalities hold:

$$\begin{aligned} \mathcal{L}(F) &= \Theta^{\blacksquare \vdash \square}(\pi_F)\,, \\ \mathcal{R}(F) &= \Theta^{\square \vdash \blacksquare}(\pi_F)\,. \end{aligned}$$

□

**Lemma 3.** For every cut-free proof $\varphi$ with end-sequent $S$, and configuration $\Omega$:

$$|\Theta^{\Omega}(\varphi)| \leq \mathcal{C}(S \cdot \Omega)\,.$$

*Remark*: $\leq$ denotes the subsumption relation, as defined in Definition 22.

*Proof.* By induction on the structure of the proof $\varphi$:

- If $\varphi$ consists just of an initial sequent $S$, then trivially:

$$|\Theta^{\Omega}(\varphi)| = \mathcal{C}(S \cdot \Omega) = S \cdot \Omega.$$

- If $\varphi$ ends with the rule $w\colon r$:

$$\dfrac{\begin{matrix}(\varphi')\\ \Gamma \vdash \Delta\end{matrix}}{\Gamma \vdash \Delta, F}\ w\colon r$$

Let $\Omega'$ be the configuration induced by $\Omega$ on $\Gamma \vdash \Delta$. By the definition of characteristic term, $\Theta^{\Omega'}(\pi') = \Theta^{\Omega}(\pi)$, and by inductive hypothesis $|\Theta^{\Omega'}(\pi')| \leq \mathcal{C}((\Gamma \vdash \Delta) \cdot \Omega')$.

- if $F$ is not tracked in $\Omega$, then $(\Gamma \vdash \Delta) \cdot \Omega' = (\Gamma \vdash \Delta, F) \cdot \Omega$ and we conclude;
- if $F$ is tracked in $\Omega$, then $\mathcal{C}((\Gamma \vdash \Delta, F) \cdot \Omega) = \mathcal{C}((\Gamma \vdash \Delta) \cdot \Omega') \times \mathcal{R}(F)$, therefore we conclude since $\mathcal{C}((\Gamma \vdash \Delta) \cdot \Omega) \leq \mathcal{C}((\Gamma \vdash \Delta, F) \cdot \Omega')$

- Similarly for the other structural rules;
- If $\varphi$ ends with the rule $\wedge\colon r$:

$$\dfrac{\begin{array}{c}(\pi_1)\\ \Gamma \vdash \Delta, F\end{array} \qquad \begin{array}{c}(\pi_2)\\ \Gamma \vdash \Delta, G\end{array}}{\Gamma \vdash \Delta, F \wedge G}\ \wedge\colon r$$

Let $\Omega_1$ and $\Omega_2$ be the configurations induced by $\Omega$ on the upper sequents $S_1$ and $S_2$, and let $S$ be the end-sequent. By inductive hypothesis, $|\Theta^{\Omega_1}(\varphi_1)| \leq \mathcal{C}(S_1 \cdot \Omega_1)$ and $|\Theta^{\Omega_2}(\varphi_2)| \leq \mathcal{C}(S_2 \cdot \Omega_2)$.

- if $F \wedge G$ is not tracked in $\Omega$, then

$$\mathcal{C}(S_1 \cdot \Omega_1) = \mathcal{C}(S_2 \cdot \Omega_2) = \mathcal{C}(S \cdot \Omega),$$

and clearly

$$|\Theta^{\Omega}(\varphi)| = |\Theta^{\Omega_1}(\varphi_1) \otimes \Theta^{\Omega_2}(\varphi_2)| \leq \mathcal{C}(S \cdot \Omega) \times \mathcal{C}(S \cdot \Omega) \leq \mathcal{C}(S \cdot \Omega).$$

- if $F \wedge G$ is tracked in $\Omega$, then for some clause set $X$, $\mathcal{C}(S_1 \cdot \Omega_1) = X \times \mathcal{R}(F)$, $\mathcal{C}(S_2 \cdot \Omega_2) = X \times \mathcal{R}(G)$, and $\mathcal{C}(S \cdot \Omega) = X \times \mathcal{R}(F \wedge G)$. We conclude by noting that

$$|\Theta^{\Omega_1}(\varphi_1) \oplus \Theta^{\Omega_2}(\varphi_2)| \leq (X \times \mathcal{R}(F)) \cup (X \times \mathcal{R}(G)) = X \times (\mathcal{R}(F) \cup \mathcal{R}(G)),$$

the last being exactly $\mathcal{R}(F \wedge G)$.

- Similarly for the other logical rules.

□

**Theorem 3.** $\mathcal{C}(S \cdot \Omega)$ is the least clause set which is subsumed by $\Theta^{\Omega}(\varphi)$ for every cut-free proof $\varphi$ with end-sequent $S$.

*Proof.* Follows from Proposition 4 and Lemma 3: by Lemma 3, $\mathcal{C}(S \cdot \Omega)$ is subsumed by every $\Theta^{\Omega}(\varphi)$, and by Proposition 4 it is itself one of those characteristic clause sets. □

**Definition 69** (Entailment)**.** By $S_1 \models S_2$ we mean that the clause set $S_1$ logically *entails* the clause set $S_2$.

In the proof for Lemma 5 we will need the following lemma:

**Lemma 4.** Let $X$, $X^\perp$, $Y_1$, and $Y_2$ be clause sets, where $X \cup X^\perp$ is unsatisfiable. Then $(X \times Y_1) \cup (X^\perp \times Y_2) \models Y_1 \times Y_2$.

*Proof.* Easy, similar as the proof for Lemma 1. □

**Lemma 5.** For every proof $\varphi$ with end-sequent $S$ and configuration $\Omega$, $|\Theta^\Omega(\varphi)|$ entails $\mathcal{C}(S \cdot \Omega)$.

*Proof.* The proof proceeds by induction on the structure of the proof $\varphi$, in the same way as in Lemma 3. The only new part is for the case of the *cut* inference:

- ...
- If $\varphi$ ends with a *cut*:

$$\dfrac{\begin{array}{c}(\pi_1)\\ \Gamma \vdash \Delta, F\end{array} \qquad \begin{array}{c}(\pi_2)\\ F, \Pi \vdash \Lambda\end{array}}{\Gamma, \Pi \vdash \Delta, \Lambda} \wedge : r$$

  Let as usual $\Omega_1$ and $\Omega_2$ be the configurations induced by $\Omega$ on the upper sequents $S_1$ and $S_2$, and let $S$ be the end-sequent. Let also $X := \mathcal{R}(F)$ and $X^\perp := \mathcal{L}(F)$, and define the clause sets $Y_1$ and $Y_2$ in such a way that $S_1 \cdot \Omega_1 = X \times Y_1$ and $S_2 \cdot \Omega_2 = X^\perp \times Y_2$. By the definition of $\mathcal{C}(\cdot)$ in Proposition 4, we have that $X \cup X^\perp = \mathcal{C}(F)$, and since it is a characteristic clause set, it is unsatisfiable. By Lemma 4, we conclude that $\Theta^\Omega(\varphi)$ entails $Y_1 \times Y_2$, and we can conclude.

□

**Theorem 4.** $\mathcal{C}(S \cdot \Omega)$ is the least clause set which is entailed by $\Theta^\Omega(\varphi)$ for every proof $\varphi$ with end-sequent $S$.

*Proof.* Follows from Proposition 4 and Lemma 5: by Lemma 5, $\mathcal{C}(S \cdot \Omega)$ is entailed by every $\Theta^\Omega(\varphi)$, and by Proposition 4 it is itself one of those characteristic clause sets. □

A consequence of Lemma 5 is the fundamental fact in Ceres that characteristic clause sets are unsatisfiable:

**Corollary 1.** CL($\varphi$) is unsatisfiable.

*Proof.* Just consider Lemma 5 with $\Omega = \varnothing$, by noting that $S \cdot \varnothing$ is $\vdash$. □

We are going to use canonic clause sets in the next chapter, in order to get characteristic clause sets which have the most general form, and use them to provide examples or counterexamples.

But the reason for introducing canonic clause sets is that they turn useful in the schematic case. Suppose we have a characteristic term schema $\Theta(\varphi)$ for a proof schema $\varphi$: if $\Theta(\varphi)$ is defined recursively, then we could refute it in the following way, by induction on $\alpha$. For the base case, we apply resolution steps to $\Theta^{\Omega}(\varphi) \downarrow 0$ for every configuration $\Omega$, obtaining $\mathcal{C}(S \downarrow 0 \cdot \Omega)$. For the inductive case, we need to resolve $\Theta^{\Omega}(\varphi) \downarrow (\alpha + 1)$ and obtain $\mathcal{C}((S \downarrow \alpha + 1) \cdot \Omega)$. Since $\Theta^{\Omega}(\varphi) \downarrow (\alpha + 1)$ depends recursively on $\Theta^{\Omega'}(\varphi) \downarrow \alpha$ for other configurations $\Omega'$, we first resolve those clause sets inside it: by inductive hypothesis, $\Theta^{\Omega'}(\varphi) \downarrow \alpha$ resolves to $\mathcal{C}((S \downarrow \alpha) \cdot \Omega')$. After applying these resolutions, we obtain a "constant" clause set which can easily resolved to $\mathcal{C}((S \downarrow \alpha) \cdot \Omega)$. Taking then $\Omega = \varnothing$ yields a refutation for $\Theta(\varphi) \downarrow (\alpha + 1)$.

CHAPTER 4

# Incompleteness in Ceres for Propositional Schemata

In this chapter, we provide our investigation of slight variations of the definitions of schemata which we introduced in Chapter 2.

Our goal is to solve the most important step of Ceres for propositional schemata: the schematic refutation of the characteristic term schema. This is a crucial step towards completeness (Theorem 1) of Ceres for propositional schemata. In order to do so, we study three restrictions to recursive definitions of resolution proof schemata:

1. the first one is in Definition 59. We allow only resolution proof schemata of the form $\rho(n)$, where $\rho$ is a resolution proof symbol, and $n$ is the arithmetic parameter: we thus disallow all the other variables in the specification of the resolution proof schema - neither arithmetic variables, nor clause variables;

2. in the second one, resolution proof schemata can have the form

$$\boldsymbol{\rho}(n; X_1, \ldots, X_\alpha)$$

where $\rho$ and $n$ are as above, and $X_1, \ldots, X_\alpha$ are variables for *clause schemata*. We disallow arithmetic variables (other than the first parameter);

3. in the third one, we allow resolution proof schemata of the most general form

$$\boldsymbol{\rho}(n; x_1, \ldots, x_\alpha; X_1, \ldots, X_\beta)$$

where $x_1, \ldots, x_\alpha$ are variables for arithmetic expressions. We show that resolution proof schemata defined by primitive recursion are not sufficient to represent schematic refutations of certain characteristic term schemata.

## 4.1 Clause variables

Let us consider in this section the form of resolution proof schemata which we introduced in Definition 59. The recursion has one integer parameter (the argument of the recursion), but we disallow any other argument. This means, no variables for integers, formulas, or clauses.

This form of schemata is *incomplete*, which means that it is not sufficient to represent refutations for all characteristic term schemata. We are going to prove this result, by first showing that every such resolution proof schema can use only a polynomial number of clauses. By providing a clause set term schema with an exponential number of clauses (all of which necessary for the refutation), we will conclude.

Let us first formalize what is a polynomial, and how a certain kind of iteration of sums of polynomials still yields a polynomial.

**Definition 70** (Polynomials in $n$)**.** $\mathbb{Q}[n]$ is the polynomial ring over the set of rational numbers, in the variable $n$.

Examples of polynomials in $\mathbb{Q}[n]$ are: 4, $n+1$, and $\frac{1}{2} \cdot n^2 + \frac{1}{4}n + \frac{1}{2}$.

**Definition 71** (Evaluation of polynomials)**.** Let $p(n) = p_0 + p_1 \cdot n + \ldots + p_\alpha \cdot n^\alpha$ be a polynomial in $\mathbb{Q}[n]$, and $N$ a natural number. We define the *evaluation* of $p(n)$ as:

$$p(n) \downarrow N := p_0 + p_1 \cdot N + \ldots + p_\alpha \cdot N^\alpha.$$

We will use polynomials to bound the number of clauses used in resolution proofs. But since we are dealing with resolution proof schemata, we need to iterate the addition of polynomials. The following lemma shows that, by iterating the sum of polynomial, we still get polynomials.

**Lemma 6.** Let $k \in \mathbb{Q}$ and $p(n) \in \mathbb{Q}[n]$; then there exists $q(n) \in \mathbb{Q}[n]$ such that:

$$\begin{aligned} q(0) &= k \\ q(n+1) &= q(n) + p(n) \end{aligned}$$

*Remark.* $q$ satisfies $q(n) = k + \sum_{i=0}^{n-1} p(i)$.

*Proof.* See for example: [Bea96]. □

Now we can use polynomials to count the clauses used in resolution proof schemata.

**Definition 72** (Number of clauses)**.** We define a bound $\#(\cdot)$ on the number of unique clauses occurring in a resolution proof schema. By structural induction on the schema $\rho$:

- if $\rho$ is a clause or a clause schema, then $\#(\rho) = 1$;

- if $\rho = \mathrm{r}(\rho_1, \rho_2; A)$, then $\#(\mathrm{r}(\rho_1, \rho_2; A)) = \#(\rho_1) + \#(\rho_2)$;
- if $\rho$ is the variable $X$, then $\#(X) = 0$;
- if $\rho = \boldsymbol{\rho}(a)$, where $\rho$ is defined by primitive recursion and $a$ is an arithmetical term

$$\begin{cases} \boldsymbol{\rho}(0) & \equiv \rho_{\text{base}}, \\ \boldsymbol{\rho}(n+1) & \equiv \rho_{\text{rec}}. \end{cases}$$

Then $\#(\boldsymbol{\rho}(a)) = \#(\boldsymbol{\rho})(a)$, where the polynomial $\#(\boldsymbol{\rho})$ is defined as in Lemma 6 by:

$$\begin{cases} \#(\boldsymbol{\rho})(0) & \equiv \#(\rho_{\text{base}}) \\ \#(\boldsymbol{\rho})(n+1) & \equiv \#(\rho_{\text{rec}}) + \#(\boldsymbol{\rho})(n) \end{cases}$$

**Theorem 5.** For every natural number $\alpha$, the number of unique clauses occurring in the resolution proof $\rho \downarrow \alpha$ is bounded by $\#(\rho) \downarrow \alpha$.

*Proof.* Follows by Definition 72. □

Since the bound $\#(\rho)$ is polynomial, resolution refutations cannot use an exponential number of clauses. But there exist characteristic clause sets that contain an exponential number of clauses, all of which are necessary in the refutation.

Consider for example the following characteristic term schema:

$$\begin{aligned} \boldsymbol{T}(0) & \equiv [\vdash P(0)] \oplus [P(0) \vdash], \\ \boldsymbol{T}(n+1) & \equiv T(n) \otimes ([\vdash P(n+1)] \oplus [P(n+1) \vdash]) \end{aligned}$$

*Remark.* $\mathbf{T}(n) \downarrow N = \mathrm{CL}^t\,(\wr P(0), \ldots, P(N) \wr)$ (top clause sets are introduced in Definition 67). $\mathbf{T}(n)$ is the characteristic term schema of a proof schema, as constructed in Example 14.

$\mathbf{T}(n) \downarrow \alpha$ contains a number of clauses which is exponential in $\alpha$. Every proper subset of that clause set is satisfiable, thus a refutation for it must necessarily use all of the clauses which it contains. We conclude that, by Theorem 5, there is no resolution refutation schema for $\mathbf{T}$, if we do not allow a less restrictive way of defining resolution proof schema.

Indeed, if we allow *clause variables* in the specification of resolution proof schemata, we can define the following refutation schema:

$$\begin{aligned} \boldsymbol{\rho}'(0 \quad ; X) & \equiv X \\ \boldsymbol{\rho}'(n+1; X) & \equiv \mathrm{r}(\boldsymbol{\rho}'(n; X \circ (\vdash P(n))), \boldsymbol{\rho}'(n; X \circ (P(n) \vdash); P(n)) \end{aligned}$$

We obtain that $\rho$ is a refutation schema for $\mathbf{T}$, where $\rho := \boldsymbol{\rho}'(n+1; \vdash)$.

In order to define the refutation above, we need to modify our definitions of clause schema and resolution proof schema.

**Definition 73** ((Modified) Clause schema)**.** The set of *modified clause schemata* is inductively defined by:

- a sequent schema which contains only atom schemata is a clause schema;
- we assume a countably infinite set of variables for clause schemata, denoted by $X_1^{\text{clause}}$, $X_2^{\text{clause}}$, ...;
- $C_1 \circ C_2$ is a clause schema, where $C_1$ and $C_2$ are clause schemata;
- $\mathbf{C}(a)$ is a clause schema, where $a$ in an arithmetical term, and $\mathbf{C}$ is defined by primitive recursion by:
$$\begin{array}{ll} \mathbf{C}(0) & \equiv \mathrm{C}_{\text{base}} \\ \mathbf{C}(n+1) & \equiv \mathrm{C}_{\text{rec}} \end{array}$$
where $\mathrm{C}_{\text{base}}$ is a clause, and $\mathrm{C}_{\text{rec}}$ is a clause schema with possible occurrences of the variable $X^{\text{clause}}$.

As we already did many times in Chapter 2, we may now define the notions of *replacement of clause schemata*, and *evaluation of clause schemata*.

**Example 15.** We can now define recursively the clause schema

$$\begin{array}{ll} \mathbf{C}(0) & \equiv P(0) \vdash \\ \mathbf{C}(n+1) & \equiv (P(n+1) \vdash) \circ \mathbf{C}(n) \end{array}$$

For every natural number $\alpha$, we have:

$$\mathbf{C}(n) \downarrow \alpha = P(0), \ldots, P(\alpha) \vdash .$$

We then need to modify Definition 59 to allow clause schemata as arguments of resolution refutation proofs:

**Definition 74** ((Modified) Resolution proof schema)**.** We define inductively the set of *resolution proof schemata* similarly as in Definition 59, with the only difference being in defined resolution proof symbols:

- a clause schema is a resolution proof schema;
- ...;
- an expression of the form $X^{\text{res}}(n; C_1, \ldots, C_\alpha)$ is a modified resolution proof schema, where $C_1$, ..., $C_\alpha$ are resolution proof schemata;

- $\boldsymbol{\rho}(a; C_1, \ldots, C_\alpha)$ is a resolution proof schema, where $a$ is an arithmetic expression, $C_1$, ..., $C_\alpha$ are resolution proof schemata, and $\boldsymbol{\rho}$ is a resolution proof schema defined by a primitive recursive specification of the form:

$$\begin{array}{lcl} \boldsymbol{\rho}(0 \quad ; X_1^{\text{clause}}, \ldots, X_\alpha^{\text{clause}}) & \equiv & \rho_{\text{base}} \\ \boldsymbol{\rho}(n+1; X_1^{\text{clause}}, \ldots, X_\alpha^{\text{clause}}) & \equiv & \rho_{\text{rec}} \end{array}$$

  where $\rho_{\text{base}}$ is a resolution proof schema, and $\rho_{\text{rec}}$ is a resolution proof schema in which can occur the resolution proof schema variable $X^{\text{res}}$. In both terms can occur the clause schema variables $X_1^{\text{clause}}$, ..., $X_\alpha^{\text{clause}}$.

And as usual, we may extend the notion of *replacement of clause schemata*, *of resolution proof schemata*, and the *evaluation of proof schemata.*

## 4.2 Integer variables

In the previous section, we allowed no integer and clause variables in recursive resolution proof schemata, and we proved that this restriction makes Ceres for propositional schemata incomplete. Then, we noted that additional expressivity is obtained by allowing clause variables in the resolution proof schemata.

Let us then consider a different restriction to resolution proof schemata: we admit clause variables, but disallow integer variables.

We will show that even this new restriction is incomplete, by providing a characteristic term schema $T$ for which no schematic refutation can be specified.

Consider first the following clause set term schemata:

$$\begin{array}{ll} |\mathbf{X}(n)| & := \{\vdash P(0); \ldots; \vdash P(n)\}, \\ |\mathbf{X}'(n)| & := \{P(0), \ldots, P(n) \vdash\} \end{array}$$

and similarly

$$\begin{array}{ll} |\mathbf{Y}(n)| & := \{\vdash Q(0); \ldots; \vdash Q(n)\}, \\ |\mathbf{Y}'(n)| & := \{Q(0), \ldots, Q(n) \vdash\}. \end{array}$$

Both the terms $\mathbf{X}(n) \oplus \mathbf{X}'(n)$ and $\mathbf{Y}(n) \oplus \mathbf{Y}'(n)$ are (informally) the characteristic term schemata of certain schematic proofs:

$$\begin{array}{ll} |\mathbf{X}(n) \oplus \mathbf{X}'(n)| & \approx \mathcal{C}\left(\bigwedge_{i=0}^n P(n)\right) \\ |\mathbf{Y}(n) \oplus \mathbf{Y}'(n)| & \approx \mathcal{C}\left(\bigwedge_{i=0}^n Q(n)\right) \end{array}$$

where $\mathcal{C}(\cdot)$ denotes the canonic characteristic clause set of a formula, as defined in Proposition 4.

It is easy to see that, because of their elementary structure, these characteristic term schemata can be refuted with the primitive recursive resolution proof schemata of Definition 59, where no integer variables are necessary.

Consider now the following clause set term schema:

$$T := (\mathbf{X}(n) \otimes \mathbf{Y}(n)) \oplus (\mathbf{X}'(n) \oplus \mathbf{Y}'(n)).$$

$T$ is a characteristic term schema too, thus every ground instance of it can be refuted; informally, it is the following canonic characteristic term schema:

$$|T \downarrow \alpha| = \mathcal{C}\left(\left(\bigwedge_{i=0}^{\alpha} P(n)\right) \vee \left(\bigwedge_{i=0}^{\alpha} Q(n)\right)\right)$$

We claim that there is no resolution refutation schema for $T$, when we restrict schemata to have no integer variables in their specification. Intuitively, in order to refute $T \downarrow \alpha$, one needs two nested resolution "routines" of length $\alpha$: but it is not possible to repeat a routine of length $n$ for $n$ times, as there is no way of keeping track of the parameter $n$ down in the recursion.

Let us make precise the notion of *depth of a resolution derivation*:

**Definition 75** (Depth of resolution derivations)**.** We define inductively the *depth* of a resolution derivation:

- For a derivation consisting of just a clause $C$, $\text{depth}(C) := 0$;
- For a derivation of the form $\text{r}(\gamma_1, \gamma_2; A)$, we set:

$$\text{depth}(\text{r}(\gamma_1, \gamma_2; A)) := 1 + \max\{\text{depth}(\gamma_1), \text{depth}(\gamma_2)\}.$$

**Lemma 7.** All refutations for $T \downarrow \alpha$ have depth $> 2 \cdot \alpha$.

*Proof.* It is easy to see, by noting that for every natural number $\alpha$, $T \downarrow \alpha$ has two clauses of length $\alpha$, and to resolve each atom in one of the two, one has to do $\alpha$ resolution steps in the other. □

First of all, let us check out what is in the characteristic term $T$. Clauses in $T \downarrow \alpha$ can be have one of the following two shapes:

1. $\vdash P(h), Q(k)$ for $h, k$ natural numbers between 0 and $\alpha$;
2. $P(0), \ldots, P(\alpha) \vdash$ or $Q(0), \ldots, P(\alpha) \vdash$.

The following lemma can be proven by reasoning by cases on the clauses in $|T \downarrow \alpha|$:

**Lemma 8.** Every proper subset of $T \downarrow \alpha$ is satisfiable. Therefore, every resolution refutation of $T \downarrow \alpha$ must use all of its clauses.

We devote the rest of this section in attempting to represent a resolution refutation schema in the formalism of the last section.

There are different ways of defining a resolution proof schema, in such a way to obtain depths bigger than $2 \cdot \alpha$:

- calling a defined resolution proof schema which executes more than one resolution step for every recursion step. For example:

$$\begin{aligned} \boldsymbol{\rho}(0) &\equiv \ldots \\ \boldsymbol{\rho}(n+1) &\equiv \mathrm{r}(\ldots, \mathrm{r}(\ldots, \ldots; \ldots); \ldots). \end{aligned}$$

- calling a defined resolution proof schema on an arithmetic expression whose coefficient for $n$ is bigger than one, for example $\boldsymbol{\rho}(2 \times n + \ldots)$.

The second option is not possible. In fact, if let's say $\rho$ calls another resolution proof schemata $\boldsymbol{\rho}'(2 \times n + \ldots)$, then for $\alpha$ big enough, $\rho \downarrow \alpha$ would call $\boldsymbol{\rho}'(2 \times n + \ldots)$, which would apply resolution steps on atom schemata whose arithmetic index is bigger than $\alpha$. But in $T \downarrow \alpha$ the atoms have indices which are natural numbers $\leq \alpha$. Therefore the only possibility is for $\boldsymbol{\rho}'$ to be a trivial resolution schema, which can be replaced by a non-recursive one.

Thus the first option should hold, but an additional requirement is necessary: in fact, any recursive resolution proof schema cannot call itself more than once per recursion step, because otherwise the resulting resolution derivations would use an exponential number of clauses, which is just not available (or, at least, the proof would just be very redundant).

Note that $|T \downarrow \alpha|$ contains $\alpha^2 + 2$ clauses: this means that the resolution proof schema should carry during every recursion step a number of resolution steps proportional to $\alpha$. This is only possible if there are *nested* resolution proof schemata, which means that we call a recursive resolution proof schema which calls another recursive resolution proof schema.

The discussion above suggests that it may be impossible to formalize our refutation schema as a resolution proof schema like the ones we introduced in the last section.

But if instead we allow the full syntax for the definition of resolution proof schemata as in [DLRW13], then schemata can have the more general form

$$\boldsymbol{\rho}(n; x_1, \ldots, x_\alpha; X_1, \ldots, X_\beta)$$

where $x_1, \ldots, x_\alpha$ are variables for arithmetic expressions and $X_1, \ldots, X_\beta$ are variables for clause schemata. Then we can specify:

$$
\begin{aligned}
\boldsymbol{\rho}_1(0; X, Z) &\equiv \mathrm{r}([\vdash P(0)] \circ Z, X \circ [P(0) \vdash]; P(0)) \\
\boldsymbol{\rho}_1(n+1; X, Z) &\equiv \mathrm{r}([\vdash P(0)] \circ Z, \boldsymbol{\rho}_1(n; X \circ [P(n+1) \vdash], Z); P(n+1)) \\
\\
\boldsymbol{\rho}_2(0; m; Y) &\equiv \mathrm{r}(\boldsymbol{\rho}_1(m; [\vdash], [\vdash Q(0)]), Y \circ [Q(0) \vdash]; Q(0)) \\
\boldsymbol{\rho}_2(n+1; m; Y) &\equiv \mathrm{r}(\boldsymbol{\rho}_1(m; [\vdash], [\vdash Q(n+1)]), \boldsymbol{\rho}_2(n; m; Y \circ [Q(0) \vdash]); Q(n+1))
\end{aligned}
$$

Finally, $\boldsymbol{\rho}_2(n, n; [\vdash])$ is a resolution refutation schema for $T$.

The difference here is that $\boldsymbol{\rho}_2$ can now store an additional integer variable $m$, which can then use to call $\boldsymbol{\rho}_1$.

Of course, this change requires to modify again the definitions we already provided for arithmetic terms, atom schemata, and the other layers of schematization. In particular, we need to introduce an infinite number of arithmetic variables. The problem is that this change creates an asymmetry between the specification of proof schemata (which use only one variable), and the specification of resolution proof schemata (which can use more arithmetic variables). But, as we will see in the next section, this change is still not sufficient to provide a complete schematic formalism for resolution refutations.

## 4.3 Primitive recursive resolution proof schemata

In the previous sections, we saw how both integer variables and clause variables are necessary in the specification of resolution proof schemata.

In this section, we are going to prove that this is still not sufficient: we are going to provide a characteristic term schema whose refutation schema is inherently not primitive recursive. This means that the schematic formalism for resolution proofs has to be radically changed with respect to the one used in Ceres for first-order schemata.

Let $A$ and $B$ two propositional atoms. Let us define the formula schemata $\mathbf{P}$ and $\mathbf{Q}$:

$$
\begin{aligned}
\mathbf{P}(0) &\equiv A, \\
\mathbf{P}(n+1) &\equiv \neg \mathbf{P}(n),
\end{aligned}
$$

and similarly:

$$
\begin{aligned}
\mathbf{Q}(0) &\equiv \neg B, \\
\mathbf{Q}(n+1) &\equiv \neg \mathbf{Q}(n).
\end{aligned}
$$

Let $T$ be the canonic characteristic term schema for the formula schema $F := \mathbf{P}(n) \wedge \mathbf{Q}(n)$, that is $|T \downarrow \alpha| := \mathcal{C}\,(F \downarrow \alpha)$.

$T$ has the specification $T = (T_P^L(n) \otimes T_Q^L(n)) \oplus (T_P^R(n) \oplus T_Q^R(n))$, where:

$$\begin{array}{llll} T_P^L(0) & \equiv [A \vdash] & T_P^R(0) & \equiv [\vdash A] \\ T_P^L(n+1) & \equiv T_P^R(n) & T_P^R(n+1) & \equiv T_P^L(n) \end{array}$$

$$\begin{array}{llll} T_Q^L(0) & \equiv [\vdash B] & T_Q^R(0) & \equiv [B \vdash] \\ T_Q^L(n+1) & \equiv T_Q^R(n) & T_Q^R(n+1) & \equiv T_Q^L(n) \end{array}$$

For every natural number $N$, $T \downarrow N$ has a very simple structure:

$$|T \downarrow N| = \begin{cases} \{\vdash B; B \vdash A; A \vdash\} & \text{if } N \text{ even,} \\ \{\vdash A; A \vdash B; B \vdash\} & \text{if } N \text{ odd.} \end{cases}$$

As we see above, $T \downarrow N$ depends on whether $N$ is even or odd; otherwise, it is constant. In addition, it is clear that the refutation of $T \downarrow N$ is straightforward, and it always has depth 2.

**Theorem 6.** Although $T \downarrow N$ is unsatisfiable for every $N$, there exists no resolution refutation schema for $T$.

*Proof.* The reason for this is that primitive recursion, at least in the form we are using, is not sufficient to specify the refutation, even though it is quite simple.

The idea of the proof is as follows: suppose $\rho$ is a resolution refutation schema for $T$. In $\rho$ there clearly need to be occurrences of resolution proof schemata defined by recursion. Now, these schemata cannot carry out resolution steps in the inductive case, otherwise the depth of the corresponding resolution deductions would increase with $N$, which is not the case.

In addition, the order of the resolution steps cannot alternate forever: for $N$ big, the resolution derivation will always either resolve first $A$ and then $B$, or vice versa. After this remark, it is easy to see that it is impossible to specify the alternation of the two resolution derivations. □

We now propose three ways to overcome the problem:

**Solution 1**. One solution would be to allow variables for *propositional symbols* or *propositional atoms* in the specification of resolution proof schemata. This would enable us to to define the following resolution proof schema:

$$\begin{array}{ll} \boldsymbol{\rho}(0 \quad ; X, Y) & \equiv \mathrm{r}(\mathrm{r}((\vdash X), (X \vdash Y); X), (Y \vdash); Y) \\ \boldsymbol{\rho}(n+1; X, Y) & \equiv \boldsymbol{\rho}(n; Y, X) \end{array}$$

A resolution refutation schema for $T$ would then be $\boldsymbol{\rho}(n; B, A)$.

**Solution 2**. Another possibility is not requiring resolution proof schemata to be defined by primitive recursion. By allowing *mutual recursion*, as in the specification of clause set term schemata, it is possible to specify the following resolution proof schema:

$$\begin{aligned}
\boldsymbol{\rho}(0) &\equiv \mathrm{r}(\mathrm{r}(\vdash B; B \vdash A; B); A \vdash; A)\\
\boldsymbol{\rho}(n+1) &\equiv \boldsymbol{\rho}'(n)\\
\boldsymbol{\rho}'(0) &\equiv \mathrm{r}(\mathrm{r}(\vdash A; A \vdash B; A); B \vdash; B)\\
\boldsymbol{\rho}'(n+1) &\equiv \boldsymbol{\rho}(n)
\end{aligned}$$

The resolution refutation schema for $T$ is then $\boldsymbol{\rho}(n)$.

**Solution 3**. The last possibility is modifying another level of the resolution proof schemata: the *arithmetic expressions*. In fact, we allowed in propositional schemata only arithmetic expressions of the form $\alpha \times n + \beta$, with $\alpha$ and $\beta$ natural numbers; and we restricted terms in schemata to have that same form.

We can revise this requirement, and permit a more expressive form of arithmetic terms: as in Ceres for first-order schemata, we may allow any function defined by a primitive recursive specification.

After this variation, we can again specify a refutation schema for the characteristic term schema above. We encode the alternating behavior of the clause set in an arithmetic function $g(n)$ such that $g(n) \downarrow N = 0$ if $N$ is even, and $g(n) \downarrow N = 1$ otherwise.

$$\begin{aligned}
f(0) &\equiv 1\\
f(n+1) &\equiv 0
\end{aligned}$$

$$\begin{aligned}
g(0) &\equiv 0\\
g(n+1) &\equiv f(g(n))
\end{aligned}$$

We can now provide a refutation schema for $T$: it is $\boldsymbol{\rho}(g(n))$, where $\boldsymbol{\rho}$ is defined by primitive recursion by:

$$\begin{aligned}
\boldsymbol{\rho}(0) &\equiv \mathrm{r}(\mathrm{r}(\vdash B; B \vdash A; B); A \vdash; A)\\
\boldsymbol{\rho}(n+1) &\equiv \mathrm{r}(\mathrm{r}(\vdash A; A \vdash B; A); B \vdash; B).
\end{aligned}$$

It is clear that, by changing the syntax of resolution proof schemata, we create an asymmetry with respect to the specification of proof schemata: defined function symbols do not occur in proof schemata, but can occur in resolution proof schemata. This difference is inelegant, thus a more conservative change of schemata is desirable. But in fact, the asymmetry is intrinsic to our problem: specifying a proof schema is necessarily easier than refuting its characteristic term schema. For example, Peano arithmetic could be expressed in a more powerful schematic language, and the specification of the refutation of schematic characteristic term schemata which is complete would yield the *consistency* of Peano arithmetic. This is impossible by well-known Gödel's *Second Incompleteness Theorem*: therefore this formalism must be more complex!

Rather than working with the different versions of resolution proof schemata which we described above, in Chapter 5 we are going to consider a totally different notion of *refutation schema*. We will prove that this notion is strong enough to parametrically represent the refutation proofs of every characteristic term schema, thus partially solving the problem of completeness.

CHAPTER 5

# Towards a Complete Ceres for Propositional Schemata

In the previous chapter, we proved the incompleteness of Ceres for propositional schemata with respect to three different flavors of resolution proof schemata.

Those results strongly suggest that if we aim at providing a complete Ceres method for propositional schemata, we need to extend the schematic language for resolution proofs in an essential way. At the end of the last chapter, we proposed a way to attempt to recover completeness: allowing arithmetic function symbols defined by primitive recursion, at least in the specification of resolution proof schemata. We noted that this change is not desirable, since it introduces an asymmetry in the notion of arithmetic term on the different levels.

In this chapter, we propose instead a slightly different version of refutations. This new method is based on *two-steps* refutation schemata:

- in the first step, a characteristic term schema is *saturated* to a top clause set, which means that weakenings are applied to the clauses in the former, in order to obtain a clause set which is in the most generic form possible;
- in the second step, the top clause set is refuted. Since this clause set term schema has a very regular structure, a resolution proof schema in the usual form suffices to represent its refutation.

Ceres for propositional schemata will prove to be complete with respect to this modified version of refutation schemata.

## 5.1 Saturation

At the basis of the method in this chapter is the fact that every unsatisfiable clause set can be reduced, by means of weakenings, to the most generic clause set, called the *top clause set*.

**Definition 76** (Atoms)**.** Let $C = A_1, \ldots, A_\alpha \vdash B_1, \ldots, B_\beta$ be a clause. We denote the atoms occurring in $C$ by Atoms($C$):

$$\mathrm{Atoms}(A_1, \ldots, A_\alpha \vdash B_1, \ldots, B_\beta) := \wr A_1, \ldots, A_\alpha, B_1, \ldots, B_\beta \int .$$

If now $S$ is a clause set, we extend the definition of Atoms($\cdot$) by

$$\mathrm{Atoms}(S) := \bigcup_{C \in S} \mathrm{Atoms}(C).$$

**Proposition 5.** Let $S$ be a clause set. Every clause in $S$ either:

- is a tautology, or
- subsumes a clause in $\mathrm{CL}^t\,(\mathrm{Atoms}(S))$.

*Proof.* Suppose that $C$ is not a tautology. Then we can apply weakenings, which result in appending atoms either to the antecedent or to the consequent of $C$. In the end we obtain $C'$, which contains every atom from Atoms($S$) on one side or the other of the clause. $C'$ clearly belongs to $\mathrm{CL}^t\,(\mathrm{Atoms}(S))$. □

**Proposition 6.** If $S$ is unsatisfiable, then every clause in $\mathrm{CL}^t\,(\mathrm{Atoms}(S))$ is subsumed by a clause in $S$.

*Proof.* Suppose a clause $C$ in $\mathrm{CL}^t\,(\mathrm{Atoms}(S))$ is not subsumed by any clause in $S$. Let us extract from $C$ a propositional interpretation $\mathcal{I}$, and let us show that $\mathcal{I}$ is a model for $S$.

$$\mathcal{I}(A) := \begin{cases} \text{true} & \text{if } A \text{ is in the antecedent of } C \\ \text{false} & \text{if } A \text{ is in the succedent of } C \end{cases} \quad \text{for every } A \in S.$$

Let us prove by contradiction that $\mathcal{I}$ is a model for $S$. Let now $D = A_1, \ldots, A_\alpha \vdash B_1, \ldots, B_\beta$ be a clause in $S$: $\mathcal{I}$ does not satisfy $D$ if and only if $\mathcal{I}(A_1) = \ldots = \mathcal{I}(A_\alpha) =$ true and $\mathcal{I}(B_1) = \ldots = \mathcal{I}(B_\beta) =$ false. By the definition of $\mathcal{I}$, this is if and only if $A_1, \ldots, A_\alpha$ are in the antecedent of $C$, and the $B_1, \ldots, B_\beta$ are in the succedent of $C$. But this means exactly that $C$ is subsumed by $D$, which contradicts our initial assumption. It follows that $\mathcal{I}$ satisfies every clause in $S$, which contradicts the hypothesis that $S$ is unsatisfiable. □

**Definition 77** (Saturation)**.** The saturation of a clause set $S$ is denoted by Saturate($S$), and it is defined by:

$$\mathrm{Saturate}(S) := \{w(C; D) \mid C \in S, D \in \mathrm{CL}^t\,(\mathrm{Atoms}(S))\}.$$

**Remark**: $w(C;D)$ is a weakening term, as defined in Definition 27.

**Example 16.** Suppose we have the clause set:

$$S := \{\vdash P(0);\ \ P(1) \vdash\}.$$

By applying the saturation, we obtain:

$$\begin{array}{rll} \text{Saturate}(S) := & \{ & w(\vdash P(0); P(0), P(1) \vdash), w(\vdash P(0); P(0) \vdash P(1)), \\ & & w(\vdash P(0); P(1) \vdash P(0)), w(\vdash P(0); \vdash P(0), P(1)), \\ & & w(P(1) \vdash; P(0), P(1) \vdash), w(P(1) \vdash; P(0) \vdash P(1)), \\ & & w(P(1) \vdash; P(1) \vdash P(0)), w(P(1) \vdash; \vdash P(0), P(1)) \\ & \}. & \end{array}$$

## 5.2 Atom set schemata

When we consider the schematic case, the operation of saturation needs to weaken the clauses to an increasing number of atoms. In fact, if we consider a clause set term schema instead of a pure clause set, its set of atoms is not constant anymore, and thus we need a way to represent schematically the atoms which occur in the clause set term schema depending on a parameter. This is why we introduce in this section the notion of *atom set schema.*

**Definition 78** (Atom set schema)**.** An *atom set schema* $\mathcal{A}$ is a finite set of pairs

$$\mathcal{A} = \{\langle P_1, a_1\rangle, \ldots, \langle P_\alpha, a_\alpha\rangle\}$$

where each first component is a propositional symbol, and each second component is an arithmetic term.

**Definition 79** (Semantics of atom set schemata)**.** For every natural number $N$, we define the evaluation of atom set schemata as

$$\mathcal{A} \downarrow N := \bigcup_{\langle P,a\rangle \in \mathcal{A}} \langle P, a\rangle \downarrow N,$$

where

$$\langle P, \alpha \times n + \beta\rangle \downarrow N := \wr P(0), P(1), \ldots, P(\alpha \cdot N + \beta) \int .$$

We say that $\mathcal{A}$ is an atom set schema for a clause set term schema $T$ if, for every natural number $N$, we have that $\text{Atoms}(T \downarrow N) \subseteq \mathcal{A} \downarrow N$.

**Example 17.** Let $\mathcal{A} = \{\langle P, n\rangle, \langle Q, 2\rangle\}$ be an atom set schema. Then:

$$\mathcal{A} \downarrow 4 = \{P(0), P(1), P(2), P(3), P(4), Q(0), Q(1), Q(2)\}.$$

**Lemma 9** (Completeness for atom set schemata)**.** For every clause set term schema $T$, there exists an atom set schema for $T$.

*Proof.* Intuitively, we need to show that we can bound linearly the number of occurrences of every propositional symbol in the clause set term schema. This can be proven by recursion on the structure of the clause set term schema $T$; we just give a sketch:

- if $T = [C]$, then construct $\mathcal{A}$ by taking, for every propositional symbol $P$ occurring in $C$, the maximum of the arithmetic terms $a$ for which $P(a)$ occurs in $C$. We say that $\max(\alpha_1 \times n + \beta_1, \alpha_2 \times n + \beta_2) := \max(\alpha_1, \alpha_2) \times n + \max(\beta_1, \beta_2)$;
- if $T$ is $T_1 \oplus T_2$ or $T_1 \otimes T_2$, then proceed recursively, taking the maximum of the arithmetic expressions occurring in the two resulting atom set schemata;
- if $T$ is $\mathbf{T}(a)$, where $\mathbf{T}$ is a defined clause set term symbol, then proceed as follows. For every propositional symbol occurring in the definition of $\mathbf{T}$, compute recursively the arithmetic bounds $b_1$ for $\mathrm{T}_{\mathrm{base}}$ and $b_2$ for $\mathrm{T}_{\mathrm{rec}}$. Let $b$ be the maximum of $b_1$ and $b_2$. Conclude with $b[n \mapsto a]$.

□

## 5.3 Top clause set schema

Following the definition of atom set schemata, we extend the notion of top clause set (Definition 67) in order to handle a schematic number of propositional atoms.

**Definition 80** (Top clause set schema)**.** Let $\mathcal{A}$ be an atom set schema. We denote the corresponding top clause set schema with $\mathrm{CL}^t(\mathcal{A})$.

**Definition 81** (Semantics of top clause set schemata)**.**

$$\mathrm{CL}^t(\mathcal{A}) \downarrow N := \mathrm{CL}^t(\mathcal{A} \downarrow N).$$

**Lemma 10.** For atom set schema $\mathcal{A}$, its relative top clause set schema $\mathrm{CL}^t(\mathcal{A})$ can be refuted schematically.

*Proof.* An explicit construction of the resolution proof schema is quite intricate to specify. Let us give the idea of the construction. Suppose $\mathcal{A}$ is $\{\langle P_1, a_1\rangle, \ldots, \langle P_\alpha, a_\alpha\rangle\}$. The resolution proof schema should resolve away all atoms in $\mathcal{A} \downarrow 0$ on the base case, while in the recursive case it should resolve all atoms in $(\mathcal{A} \downarrow N+1) \setminus (\mathcal{A} \downarrow N)$.

It is interesting to note that the specification of the resulting resolution proof schema does not use arithmetic variables, and it uses only one clause variable.

For an example of the resolution proof schema, see the example below. □

**Example 18.** Let us take for example the simple atom set schema $\mathcal{A}$ such that:

$$\mathcal{A} \downarrow N := \{P(0), \ldots, P(N)\}.$$

Let us consider its relative top clause set schema $\mathrm{CL}^t(\mathcal{A})$. We define the resolution proof schema $\rho := \boldsymbol{\rho}'(n; \vdash)$, where $\boldsymbol{\rho}'$ is:

$$\begin{aligned}
\boldsymbol{\rho}'(0; C) &\equiv \mathrm{r}(C \circ (\vdash P(0)), C \circ (P(0) \vdash); P(0)),\\
\boldsymbol{\rho}'(n+1; C) &\equiv \mathrm{r}(\\
&\quad \boldsymbol{\rho}'(n; C \circ (\vdash P(n+1))),\\
&\quad \boldsymbol{\rho}'(n; C \circ (P(n+1) \vdash));\\
&\quad P(n+1)\\
&).
\end{aligned}$$

$\rho$ is a resolution proof schema from $\mathrm{CL}^t(\mathcal{A})$ with end-clause $\vdash$.

## 5.4 Refutation schemata

We are now ready to change the notion of refutation schemata for clause set term schemata. It cannot consist simply of a resolution proof schema - as in the usual formulation of Ceres for first-order schemata - but it should take into account the initial saturation phase.

**Definition 82** (Refutation schema)**.** A *refutation schema* for $\Theta$ is a triple of the form $\langle \Theta, \mathcal{A}, \rho \rangle$, where $\Theta$ is a characteristic term schema, $\mathcal{A}$ is an atom set schema, and $\rho$ is a resolution proof schema, such that:

- $\mathcal{A}$ is an atom set schema for $\Theta$;
- $\rho$ is a resolution deduction schema from $\mathrm{CL}^t(\mathcal{A})$ with end-clause $\vdash$.

Let $\langle \Theta, \mathcal{A}, \rho \rangle$ be a refutation schema for $\Theta$. The idea of this two-steps method is the following: first, we $w$-resolve the characteristic term schema $\Theta$ (by saturation) to the top clause set schema $\mathrm{CL}^t(\mathcal{A})$; second, we apply the resolution proof schema $\rho$ and resolve $\mathrm{CL}^t(\mathcal{A})$ to the empty clause $\vdash$.

This procedure is condensed in the following definition:

**Definition 83** (Semantics of refutation schemata)**.** $\langle \Theta, \mathcal{A}, \rho \rangle \downarrow N$ is a resolution deduction obtained from $\rho \downarrow N$ by replacing every clause $D$ which is a leaf in the deduction tree with a weakening term $w(C; D)$ such that $C \in (\Theta \downarrow N)$ and $C \leq D$.

Remark: in the definition above, the leaves in the deduction tree of $\rho \downarrow N$ are clauses in $\mathrm{CL}^t(\mathcal{A})$. Since $\Theta$ is unsatisfiable, we know by Proposition 6 that such a clause $C$ always exists.

**Theorem 7** (Soundness for refutation schemata)**.** Refutation schemata are sound, i.e. if $R = \langle \Theta, \mathcal{A}, \rho \rangle$ is a refutation schema and $N$ a natural number, then $R \downarrow N$ is a $w$-resolution deduction from $\Theta \downarrow N$.

*Proof.* Immediate by Definition 83. □

**Theorem 8** (Completeness for refutation schemata in Ceres)**.** Every characteristic clause term schema Θ has a refutation schema.

*Proof.* Follows from Lemma 9 and Lemma 10. □

CHAPTER 6

# Example

As a test case for our method, let us consider a proof schema, compute its characteristic term schema, and give a refutation schema for it.

The proof schema we are going to study formalizes the following fact: if $P(0)$ holds, and if we have the chain of implications $P(0) \supset P(1)$, $P(1) \supset P(2)$, …, $P(N) \supset P(N+1)$ of length $N$, then also $P(N+1)$ holds. We will denote this proof schema by $\boldsymbol{\psi}(n)$, with end-sequent:

$$P(0), \bigwedge_{i=0}^{n} P(i) \supset P(i+1) \vdash P(n+1).$$

First, we define a formula schema which implements the generalized conjunction.

**Formula schema: $\boldsymbol{Q}$**

$$\begin{aligned} \mathbf{Q}(0) &\equiv P(0) \supset P(1) \\ \mathbf{Q}(n+1) &\equiv \mathbf{Q}(n) \wedge (P(n+1) \supset P(n+2)) \end{aligned}$$

**Remark**: $\mathbf{Q}(n)$ defines $\bigwedge_{i=0}^{n} P(i) \supset P(i+1)$.

Second, we give the proof schema $\boldsymbol{\psi}(n)$ with end-sequent $\mathbf{Q}(n), P(0) \vdash P(n+1)$, which we specify by primitive recursion; this proof schema depends on other proof schemata, namely $\boldsymbol{\tau}$, $\boldsymbol{\chi}$, $\boldsymbol{\sigma}$, $\boldsymbol{\omega}$ and $\boldsymbol{\lambda}$.

**Proof schema: $\boldsymbol{\psi}$**

$$\boldsymbol{\psi}(0) \quad \equiv \quad \dfrac{\dfrac{P(0) \vdash P(0) \qquad P(1) \vdash P(1)}{P(0) \supset P(1), P(0) \vdash P(1)}\ \supset\colon l}{\mathbf{Q}(0), P(0) \vdash P(1)}\ \mathit{def}$$

$$\boldsymbol{\psi}(n+1) \quad \equiv \quad \dfrac{\dfrac{(\boldsymbol{\chi}(n+1))}{\mathbf{Q}(n+1) \vdash P(0) \supset P(n+2)} \qquad \dfrac{(\boldsymbol{\tau}(n+1))}{P(0), P(0) \supset P(n+2) \vdash P(n+2)}}{\mathbf{Q}(n+1), P(0) \vdash P(n+2)}\ \mathit{cut} \tag{6.1}$$

**Proof schema: $\boldsymbol{\tau}$**

$$\boldsymbol{\tau}(k) \equiv \dfrac{P(0) \vdash P(0) \qquad P(k+1) \vdash P(k+1)}{P(0), P(0) \supset P(k+1) \vdash P(k+1)}\ \supset\colon l$$

**Proof schema: $\boldsymbol{\chi}$**

$$\boldsymbol{\chi}(0) \quad \equiv \quad \dfrac{\dfrac{\dfrac{P(0) \vdash P(0) \qquad P(1) \vdash P(1)}{P(0) \supset P(1), P(0) \vdash P(1)}\ \supset\colon l}{\mathbf{Q}(0), P(0) \vdash P(1)}\ \mathit{def}}{\mathbf{Q}(0) \vdash P(0) \supset P(1)}\ \supset\colon r \tag{6.2}$$

$$\boldsymbol{\chi}(n+1) \quad \equiv \quad \text{see below} \downarrow$$

$$\dfrac{\dfrac{\dfrac{(\boldsymbol{\lambda}(n))}{\mathbf{Q}(n+1) \vdash \mathbf{Q}(n)} \quad \mathit{cut}\ \dfrac{\dfrac{\dfrac{\boldsymbol{\chi}(n)}{\mathbf{Q}(n) \vdash P(0) \supset P(n+1)} \quad \dfrac{(\boldsymbol{\sigma}(n))}{\mathbf{Q}(n+1) \vdash P(n+1) \supset P(n+2)}}{\mathbf{Q}(n), \mathbf{Q}(n+1) \vdash (P(0) \supset P(n+1)) \wedge (P(n+1) \supset P(n+2))}\ \wedge\colon r \qquad \boldsymbol{\omega}(n+1)}{\mathbf{Q}(n), \mathbf{Q}(n+1) \vdash P(0) \supset P(n+2)}}{\mathbf{Q}(n+1), \mathbf{Q}(n+1) \vdash P(0) \supset P(n+2)}\ \mathit{cut}}{\mathbf{Q}(n+1) \vdash P(0) \supset P(n+2)}\ c\colon l$$

**Proof schema: $\boldsymbol{\omega}$**

$$\boldsymbol{\omega}(n) \equiv \dfrac{\dfrac{\dfrac{\dfrac{P(0) \vdash P(0) \qquad \dfrac{P(n) \vdash P(n) \qquad P(n+1) \vdash P(n+1)}{P(n), P(n) \supset P(n+1) \vdash P(n+1)}\ \supset\colon l}{P(0), P(0) \supset P(n), P(n) \supset P(n+1) \vdash P(n)}}{P(0), (P(0) \supset P(n)) \wedge (P(n) \supset P(n+1)), P(n) \supset P(n+1) \vdash P(0) \supset P(n+1)}\ \wedge\colon l_1}{P(0), 2 \times (P(0) \supset P(n)) \wedge (P(n) \supset P(n+1)) \vdash P(0) \supset P(n+1)}\ \wedge\colon l_2}{(P(0) \supset P(n)) \wedge (P(n) \supset P(n+1)) \vdash P(0) \supset P(n+1)}\ c\colon l$$

**Proof schema: $\boldsymbol{\sigma}$**

$$\boldsymbol{\sigma}(n) \equiv \dfrac{\dfrac{\dfrac{\dfrac{P(n+1) \vdash P(n+1) \qquad P(n+2) \vdash P(n+2)}{P(n+1), P(n+1) \supset P(n+2) \vdash P(n+2)}\supset: l}{P(n+1) \supset P(n+2) \vdash P(n+1) \supset P(n+2)}\supset: r}{\boldsymbol{Q}(n) \wedge (P(n+1) \supset P(n+2)) \vdash P(n+1) \supset P(n+2)}\wedge: l_2}{\boldsymbol{Q}(n+1) \vdash P(n+1) \supset P(n+2)}\,def$$

**Proof schema: $\boldsymbol{\lambda}$**

$$\boldsymbol{\lambda}(n) \equiv \dfrac{\dfrac{\dfrac{\vdots}{\boldsymbol{Q}(n) \vdash \boldsymbol{Q}(n)}}{\boldsymbol{Q}(n) \wedge (P(n+1) \supset P(n+2)) \vdash \boldsymbol{Q}(n)}\wedge: l_1}{\boldsymbol{Q}(n+1) \vdash \boldsymbol{Q}(n)}\,def$$

In figure Figure 6.1, one can find the "call structure" for the proof schemata we defined above. We note that $\boldsymbol{\psi}$ uses the schemata $\boldsymbol{\tau}$ and $\boldsymbol{\chi}$; the latter is specified by primitive recursion, and it depends on $\boldsymbol{\sigma}$, $\boldsymbol{\lambda}$ and $\boldsymbol{\omega}$.

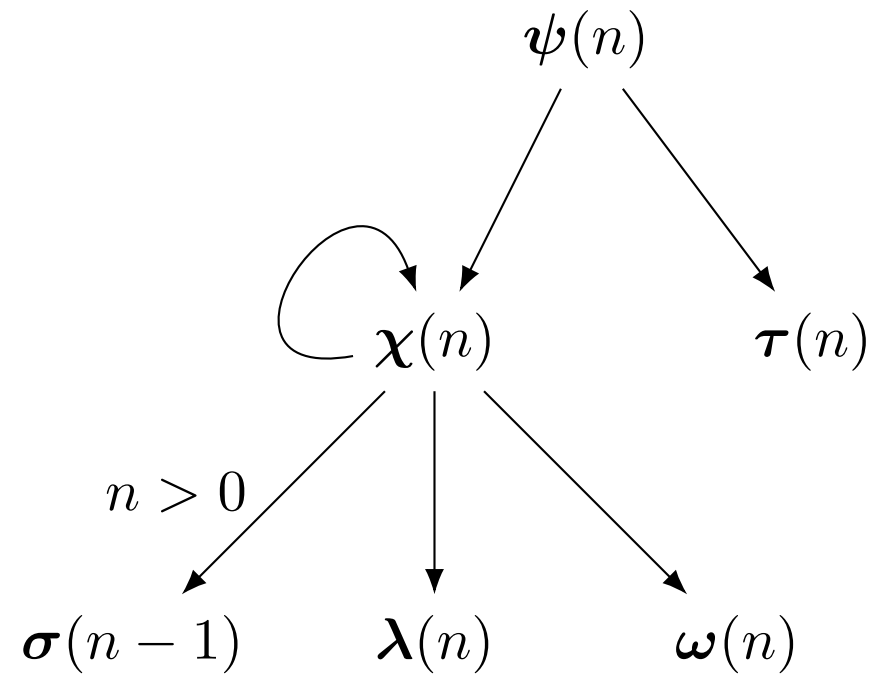

Figure 6.1: Call structure of the proof schemata

Let us now compute the characteristic term schema $\Theta\ (\boldsymbol{\psi})$. In order to do so, we start by trying to define $\Theta^{(\square,\ \square \vdash \square)}\ (\boldsymbol{\psi})$, and then we check which other characteristic clause set schemata we need to call, and on which configurations.

The definitions can be found in Figure 6.2.

We note that all the definitions of clause set term schemata in the figure are primitive recursive, i.e. it is not necessary to specify characteristic term schemata by *mutual recursion*. And in fact this is clear by looking at the dependency graph for the characteristic term schemata in Figure 6.3: the graph has no non-trivial cycles.

$$\begin{aligned}
\Theta^{(\square,\, \square \vdash \square)}\,(\boldsymbol{\psi})\,(0) &\equiv [\vdash] \otimes [\vdash] \\
\Theta^{(\square,\, \square \vdash \square)}\,(\boldsymbol{\psi})\,(n+1) &\equiv \Theta^{(\square \vdash \blacksquare)}\,(\boldsymbol{\chi})\,(n+1) \oplus \Theta^{(\square, \blacksquare \vdash \square)}\,(\boldsymbol{\tau})\,(n+1)
\end{aligned}$$

$$\Theta^{(\square, \blacksquare \vdash \square)}\,(\boldsymbol{\tau})\,(n) := [\vdash P(0)] \oplus [P(n+1) \vdash]$$

$$\begin{aligned}
\Theta^{(\square \vdash \blacksquare)}\,(\boldsymbol{\chi})\,(0) &\equiv [P(0) \vdash] \otimes [\vdash P(1)] \\
\Theta^{(\square \vdash \blacksquare)}\,(\boldsymbol{\chi})\,(n+1) &\equiv \Theta^{(\square \vdash \blacksquare)}\,(\boldsymbol{\lambda})\,(n) \oplus \Theta^{(\blacksquare \vdash \blacksquare)}\,(\boldsymbol{\chi})\,(n) \oplus \ldots \\
&\quad \ldots \oplus \Theta^{(\square \vdash \blacksquare)}\,(\boldsymbol{\sigma})\,(n) \oplus \Theta^{(\blacksquare \vdash \blacksquare)}\,(\boldsymbol{\omega})\,(n+1)
\end{aligned}$$

$$\begin{aligned}
\Theta^{(\blacksquare \vdash \blacksquare)}\,(\boldsymbol{\chi})\,(0) &\equiv [P(0) \vdash P(0)] \oplus [P(1) \vdash P(1)] \\
\Theta^{(\blacksquare \vdash \blacksquare)}\,(\boldsymbol{\chi})\,(n+1) &\equiv \Theta^{(\blacksquare \vdash \blacksquare)}\,(\boldsymbol{\lambda})\,(n) \oplus \Theta^{(\blacksquare \vdash \blacksquare)}\,(\boldsymbol{\chi})\,(n) \oplus \ldots \\
&\quad \ldots \oplus \Theta^{(\blacksquare \vdash \blacksquare)}\,(\boldsymbol{\sigma})\,(n) \oplus \Theta^{(\blacksquare \vdash \blacksquare)}\,(\boldsymbol{\omega})\,(n+1)
\end{aligned}$$

$$\Theta^{(\blacksquare \vdash \blacksquare)}\,(\boldsymbol{\omega})\,(n) := [P(0) \vdash P(0)] \oplus ([P(n) \vdash P(n)] \oplus [P(n+1) \vdash P(n+1)])$$

$$\begin{aligned}
\Theta^{(\square \vdash \blacksquare)}\,(\boldsymbol{\sigma})\,(n) &\equiv [P(n+1) \vdash] \otimes [\vdash P(n+2)] \\
\Theta^{(\blacksquare \vdash \blacksquare)}\,(\boldsymbol{\sigma})\,(n) &\equiv [P(n+1) \vdash P(n+1)] \oplus [P(n+2) \vdash P(n+2)]
\end{aligned}$$

$$\begin{aligned}
\Theta^{(\square \vdash \blacksquare)}\,(\boldsymbol{\lambda})\,(0) &\equiv [P(0) \vdash] \otimes [\vdash P(1)] \\
\Theta^{(\square \vdash \blacksquare)}\,(\boldsymbol{\lambda})\,(n+1) &\equiv \Theta^{(\square \vdash \blacksquare)}\,(\boldsymbol{\lambda})\,(n) \oplus ([P(n) \vdash] \otimes [\vdash P(n+1)])
\end{aligned}$$

$$\begin{aligned}
\Theta^{(\blacksquare \vdash \blacksquare)}\,(\boldsymbol{\lambda})\,(0) &\equiv [P(0) \vdash P(0)] \oplus [P(1) \vdash P(1)] \\
\Theta^{(\blacksquare \vdash \blacksquare)}\,(\boldsymbol{\lambda})\,(n+1) &\equiv \Theta^{(\blacksquare \vdash \blacksquare)}\,(\boldsymbol{\lambda})\,(n) \oplus [P(n) \vdash P(n)] \oplus [P(n+1) \vdash P(n+1)]
\end{aligned}$$

Figure 6.2: Useful clause set term schemata

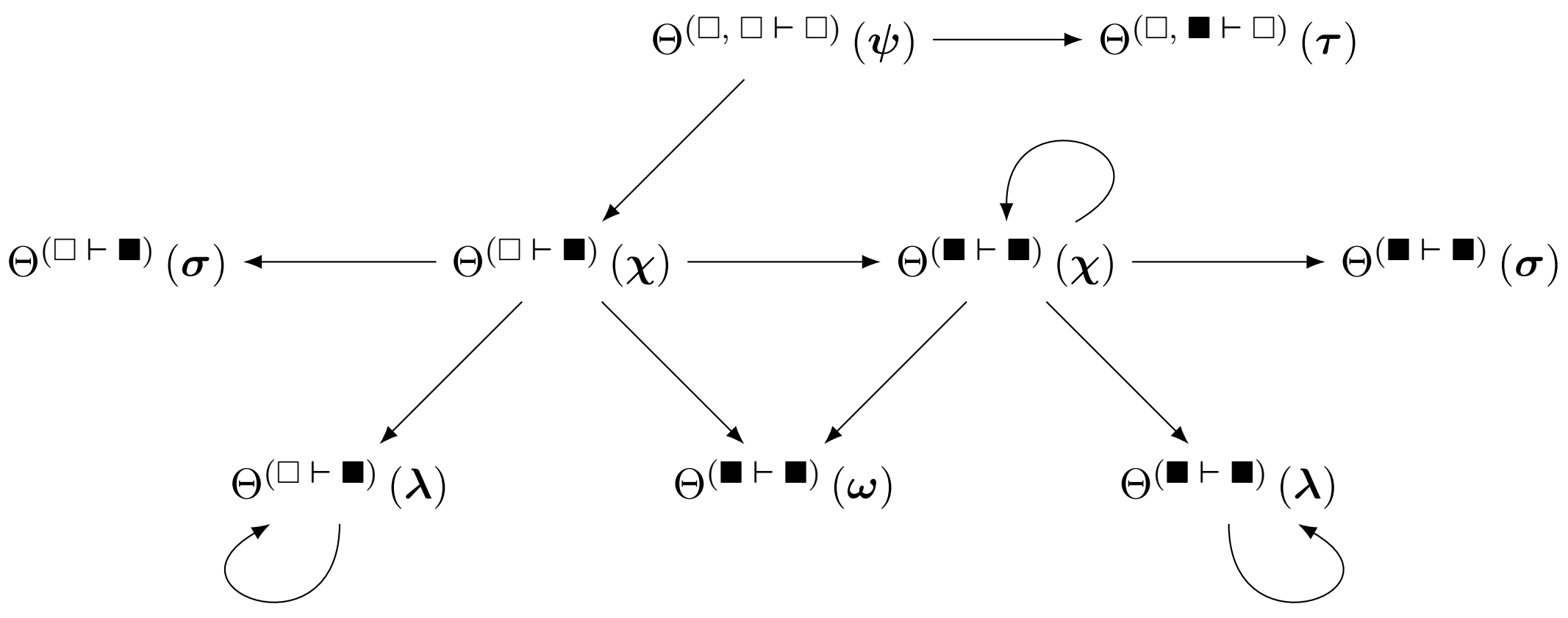

Figure 6.3: Call structure of the characteristic term schemata

| $N$ | $\lvert\Theta(\boldsymbol{\psi})(n)\downarrow N\rvert$ |
|---|---|
| 0 | $\vdash$ |
| 1 | $\vdash P(2)$; $P(0)\vdash$; $P(0)\vdash P(0)$; $P(1)\vdash P(0)$; $P(1)\vdash P(1)$; $P(2)\vdash P(1)$; $P(2)\vdash P(2)$ |
| 2 | $\vdash P(3)$; $P(0)\vdash$; $P(0)\vdash P(0)$; $P(1)\vdash P(0)$; $P(1)\vdash P(1)$; $P(2)\vdash P(1)$; $P(2)\vdash P(2)$; $P(3)\vdash P(2)$; $P(3)\vdash P(3)$ |
| 3 | $\vdash P(4)$; $P(0)\vdash$; $P(0)\vdash P(0)$; $P(1)\vdash P(0)$; $P(1)\vdash P(1)$; $P(2)\vdash P(1)$; $P(2)\vdash P(2)$; $P(3)\vdash P(2)$; $P(3)\vdash P(3)$; $P(4)\vdash P(3)$; $P(4)\vdash P(4)$ |
| 4 | $\vdash P(5)$; $P(0)\vdash$; $P(0)\vdash P(0)$; $P(1)\vdash P(0)$; $P(1)\vdash P(1)$; $P(2)\vdash P(1)$; $P(2)\vdash P(2)$; $P(3)\vdash P(2)$; $P(3)\vdash P(3)$; $P(4)\vdash P(3)$; $P(4)\vdash P(4)$; $P(5)\vdash P(4)$; $P(5)\vdash P(5)$ |
| 5 | $\vdash P(6)$; $P(0)\vdash$; $P(0)\vdash P(0)$; $P(1)\vdash P(0)$; $P(1)\vdash P(1)$; $P(2)\vdash P(1)$; $P(2)\vdash P(2)$; $P(3)\vdash P(2)$; $P(3)\vdash P(3)$; $P(4)\vdash P(3)$; $P(4)\vdash P(4)$; $P(5)\vdash P(4)$; $P(5)\vdash P(5)$; $P(6)\vdash P(5)$; $P(6)\vdash P(6)$ |
| 6 | $\vdash P(7)$; $P(0)\vdash$; $P(0)\vdash P(0)$; $P(1)\vdash P(0)$; $P(1)\vdash P(1)$; $P(2)\vdash P(1)$; $P(2)\vdash P(2)$; $P(3)\vdash P(2)$; $P(3)\vdash P(3)$; $P(4)\vdash P(3)$; $P(4)\vdash P(4)$; $P(5)\vdash P(4)$; $P(5)\vdash P(5)$; $P(6)\vdash P(5)$; $P(6)\vdash P(6)$; $P(7)\vdash P(6)$; $P(7)\vdash P(7)$ |
| 7 | $\vdash P(8)$; $P(0)\vdash$; $P(0)\vdash P(0)$; $P(1)\vdash P(0)$; $P(1)\vdash P(1)$; $P(2)\vdash P(1)$; $P(2)\vdash P(2)$; $P(3)\vdash P(2)$; $P(3)\vdash P(3)$; $P(4)\vdash P(3)$; $P(4)\vdash P(4)$; $P(5)\vdash P(4)$; $P(5)\vdash P(5)$; $P(6)\vdash P(5)$; $P(6)\vdash P(6)$; $P(7)\vdash P(6)$; $P(7)\vdash P(7)$; $P(8)\vdash P(7)$; $P(8)\vdash P(8)$ |
| 8 | $\vdash P(9)$; $P(0)\vdash$; $P(0)\vdash P(0)$; $P(1)\vdash P(0)$; $P(1)\vdash P(1)$; $P(2)\vdash P(1)$; $P(2)\vdash P(2)$; $P(3)\vdash P(2)$; $P(3)\vdash P(3)$; $P(4)\vdash P(3)$; $P(4)\vdash P(4)$; $P(5)\vdash P(4)$; $P(5)\vdash P(5)$; $P(6)\vdash P(5)$; $P(6)\vdash P(6)$; $P(7)\vdash P(6)$; $P(7)\vdash P(7)$; $P(8)\vdash P(7)$; $P(8)\vdash P(8)$; $P(9)\vdash P(8)$; $P(9)\vdash P(9)$ |
| 9 | $\vdash P(1)$; $P(0)\vdash$; $P(0)\vdash P(0)$; $P(1)\vdash P(0)$; $P(1)\vdash P(1)$; $P(1)\vdash P(9)$; $P(2)\vdash P(1)$; $P(2)\vdash P(2)$; $P(3)\vdash P(2)$; $P(3)\vdash P(3)$; $P(4)\vdash P(3)$; $P(4)\vdash P(4)$; $P(5)\vdash P(4)$; $P(5)\vdash P(5)$; $P(6)\vdash P(5)$; $P(6)\vdash P(6)$; $P(7)\vdash P(6)$; $P(7)\vdash P(7)$; $P(8)\vdash P(7)$; $P(8)\vdash P(8)$; $P(9)\vdash P(8)$; $P(9)\vdash P(9)$ |
| $\vdots$ | |

Figure 6.4: Characteristic clause sets for $\boldsymbol{\psi}(n)$

After defining the characteristic term schema for $\Theta(\boldsymbol{\psi})$, we are ready to provide a schematic refutation for it.

But first, in order to get an idea of the clauses in the characteristic clause set, let us have a look at the sets corresponding to $\Theta(\boldsymbol{\psi})\downarrow N$ for different natural numbers $N$ (Figure 6.4).

By studying the table in Figure 6.4 we note that, disregarding the tautologies (which are irrelevant), the characteristic clause term contains, for $N > 0$:

$$\Theta(\boldsymbol{\psi}) \downarrow N \supseteq \{\vdash P(N+1); P(N+1) \vdash P(N); \ldots; P(1) \vdash P(0); P(0) \vdash\}.$$

There is a simple resolution refutation from these clauses:

$$\cfrac{\cfrac{\cfrac{\vdash P(N+1) \qquad P(N+1) \vdash P(N)}{\vdash P(N)} \qquad P(N) \vdash P(N-1)}{\ddots\ \ddots\ \ddots} }{\cfrac{\vdash P(0) \qquad\qquad P(0) \vdash}{\vdash}}$$

This refutation can be specified as a resolution proof schema in the following way:

$$\begin{aligned}
\boldsymbol{\rho}'(0) &\equiv P(0) \vdash \\
\boldsymbol{\rho}'(n+1) &\equiv \mathrm{r}(P(n+1) \vdash P(n);\ \boldsymbol{\rho}'(n); P(n)) \\
\\
\boldsymbol{\rho}(0) &\equiv\ \vdash \\
\boldsymbol{\rho}(n+1) &\equiv \mathrm{r}((\vdash P(n+1)), \boldsymbol{\rho}'(n+1); P(n+1))
\end{aligned}$$

We have that $\boldsymbol{\rho}$ is a resolution proof schema from $\Theta(\boldsymbol{\psi})$, with end-clause $\vdash$.

We can conclude that, since the shape of characteristic term schema was regular enough, we were able to find a resolution proof schema for $\Theta(\boldsymbol{\psi})$ which is a refutation. But let us now use instead the new kind of refutation schemata that we introduced in the last chapter.

A refutation schema for $\Theta(\boldsymbol{\psi})$ is $\langle \Theta(\boldsymbol{\psi}), \mathcal{A}, \rho \rangle$, where $\mathcal{A}$ is an atom set schema and $\rho$ is a resolution proof schema from $\mathrm{CL}^t(\mathcal{A})$ with end-clause $\vdash$.

### Atom set schema

$|\Theta(\boldsymbol{\psi})(n) \downarrow 0| = \{\vdash\}$ does not contain occurrences of propositional atoms, and $|\Theta(\boldsymbol{\psi})(n) \downarrow\ \alpha|$ for $\alpha > 0$ contains occurrences of $P(0), \ldots, P(\alpha), P(\alpha+1)$. Thus we use the atom set schema $\mathcal{A} := \{\langle P, n+1 \rangle\}$, which yields:

$$\mathcal{A} \downarrow \alpha := \{P(0), \ldots, P(\alpha), P(\alpha+1)\}.$$

### Refutation schema

$\rho$ needs to be a resolution proof schema for the top clause set schema $\mathrm{CL}^t(\mathcal{A})$. We set $\rho := \boldsymbol{\rho}'(n+1; \vdash)$ where $\boldsymbol{\rho}'$ is:

$$\begin{aligned}
\boldsymbol{\rho}'(0; C) &\equiv \mathrm{r}(C \circ (\vdash P(0)), C \circ (P(0) \vdash); P(0)) \\
\boldsymbol{\rho}'(n+1; C) &\equiv \mathrm{r}(\boldsymbol{\rho}'(n; C \circ (\vdash P(n+1))), \boldsymbol{\rho}'(n; C \circ (P(n+1) \vdash)); P(n+1))
\end{aligned}$$

We successfully defined the refutation schema for $\Theta\left(\boldsymbol{\psi}\right)$.

As an example of the redundancy of the resolution proof schema, let us consider the w-resolution derivation $\langle \Theta\left(\boldsymbol{\psi}\right), \mathcal{A}, \rho \rangle \downarrow 0$:

$$\dfrac{\dfrac{\dfrac{\vdash}{\vdash P(0), P(1)}\,w \qquad \dfrac{\vdash}{P(0) \vdash P(1)}\,w}{\vdash P(1)} \qquad \dfrac{\dfrac{\vdash}{P(1) \vdash P(0)}\,w \qquad \dfrac{\vdash}{P(0), P(1) \vdash}\,w}{P(1) \vdash}}{\vdash}$$

CHAPTER 7

# Conclusion

This thesis was set out to answer the question whether propositional schemata could provide a simple and neat setting on which schematic Ceres would result complete.

As we saw in Chapter 4, the naive restriction of Ceres for first-order schemata to the propositional case with linear arithmetic expressions and one arithmetic variable does not work. Even modifying the syntax for the specification of refutations, making it more and more similar to the case of first-order schemata, was not sufficient.

Therefore, we followed a different method in Chapter 5. In order to refute schematically a characteristic term schema, we introduced a refutation in two steps. First, weakenings are applied to the clauses in the term schema, hence obtaining a very redundant (in the sense of subsumption) clause set, called *top clause set* (see Definition 67). The top clause set has a very uniform structure, thus it can be easily refuted with the old notion of resolution proof schemata.

Consequently, we can conclude that we provided a complete method for uniformly and schematically refuting characteristic term schemata of propositional proof schemata. This is a first step towards a complete Ceres for propositional schemata, and it could be in principle applied successfully the case of first-order schemata.

### Further work

We achieved our goal of providing a complete syntax for specifying refutation schemata of characteristic term schemata for propositional schemata. As we already noted, our method is based on a specification of refutations which is not purely primitive recursive, in the sense that a preliminary step is necessary: a characteristic term schema should first be saturated, by applying weakenings, to a top clause set schema. This step is carried outside the usual framework for resolution proof schemata in [DLRW13]. We believe that the schematic resolution calculus in [DLRW13] is inherently incomplete with respect to characteristic term schemata, and thus a radical change is in any case necessary.

Other ways of obtaining completeness for the resolution step of schematic Ceres could be:

- by a *schematic subsumption principle.* Subsumption of clause terms can be extended to subsumption of resolution proofs as in [BL11]. It would be clearly helpful to make subsumption schematic, and to study cut reduction in schemata under this new schematic subsumption principle.
- by using the *generic clause sets* specified in Chapter 3, and in particular in Section 3.2. In fact, canonic characteristic clause sets provide uniform shapes for characteristic clause sets with respect to cut configurations. It should be possible to define resolution proof schemata in a recursive way, such that at every step of the recursion, the characteristic term schema on configuration $\Omega$ is reduced to the relative canonic clause set. In this way, one could easily get a refutation schema by schematizing Corollary 1.

Another problem to be tackled, is finding a schematic way to merge refutation schemata and projection schemata. This should result in a schema of ACNF, which stands for *Atomic Cut Normal Form*: it is the result of cut-elimination by Ceres, a proof schema which contains cut inferences only on atomic formulas.